\documentclass[a4paper,11pt]{article}
\usepackage{jheppub}
\usepackage{graphicx}
\makeatletter
\renewcommand*{\thesection}{\arabic{section}}

\renewcommand*{\p@subsection}{}

\renewcommand*{\p@subsubsection}{}
\makeatother

\usepackage{amsmath}
\usepackage{pstricks}
\usepackage{color}
\usepackage{xcolor}
\usepackage{graphicx}%
\usepackage{dcolumn}%
\usepackage{bm}
\usepackage{slashed}
\usepackage{subcaption}
\usepackage{float}
\usepackage{hyperref}
\usepackage{caption}
\usepackage{threeparttable}
\usepackage{jheppub}

\newcommand{\bea}{\begin{eqnarray}}
\newcommand{\eea}{\end{eqnarray}}
\newcommand{\beq}{\begin{eqnarray}}
\newcommand{\eeq}{\end{eqnarray}}

\def\bit{\begin{itemize}
 }
 \def\eit{\end{itemize}
 }

\begin{document}

\preprint{}

\title{\centering Dark Freeze-out Cogenesis}

\author[a]{Xiaoyong Chu,}
 \emailAdd{xiaoyong.chu@oeaw.ac.at}
\author[b]{Yanou Cui,} 
 \emailAdd{yanou.cui@ucr.edu}
\author[a]{Josef Pradler,}
 \emailAdd{josef.pradler@oeaw.ac.at}
\author[c]{Michael Shamma}
 \emailAdd{mshamma@triumf.ca}
\affiliation[a]{Institute of High Energy Physics, Austrian Academy of Sciences, Nikolsdorfergasse 18, 1050 Vienna, Austria}
\affiliation[b]{Department of Physics and Astronomy, University of California, Riverside, CA 92521, USA}
\affiliation[c]{TRIUMF, 4004 Wesbrook Mall, Vancouver, BC V6T 2A3, Canada}

\date{\today}%

\abstract{
We propose a new mechanism where a multi-component dark sector generates the observed dark matter abundance and baryon asymmetry and thus addresses the coincidence between the two. The thermal freeze-out of dark matter annihilating into meta-stable dark partners sets the dark matter relic abundance while providing the out-of-equilibrium condition for baryogenesis. The meta-stable state triggers baryon asymmetry production by its decay well after the freeze-out and potentially induces a period of early matter domination before its decay. The dark matter and baryon abundances are related through number conservation within the dark sector (cogenesis). The ``coincidence" is a natural outcome with GeV- to TeV-scale symmetric dark matter and the dark sector's interactions with the Standard Model quarks. We present a UV-complete model and explore its phenomenological predictions, including dark matter direct detection signals, LHC signatures of new massive particles with color charges and long-lived particles with displaced vertices, dark matter-induced nucleon conversions, (exotic) dark matter indirect detection signals, and effects on the cosmological matter power spectrum. As a side result, we provide a novel analytical treatment for dark sector freeze-out, which may prove useful in the study of related scenarios. 
}

\maketitle
\thispagestyle{empty}
\newpage

\section{Introduction}\label{sec:intro}
\setcounter{page}{1}

The nature and dynamical origins of the baryon asymmetry and the observed dark matter (DM) are two long-standing puzzles in particle cosmology. Meanwhile, the observation that their abundances are strikingly similar, $\Omega_{DM}/\Omega_{B}\approx5$ \cite{planck}, presents a coincidence problem which is suggestive of a potential connection between the origins of DM and baryons in the early Universe. Although there exists a cornucopia of theoretical explanations which address each of these pieces separately, this cosmic coincidence has inspired new directions in DM model-building (see e.g.,~\cite{Boucenna:2013wba,Cui:2015eba} for reviews).

Weakly interacting massive particles (WIMPs) as DM candidates have been leading the DM model-building efforts since the 1990s~\cite{Primack:1988zm,Jungman:1995df}. This paradigm is motivated largely by the observation that DM with weak-scale interactions and masses can produce the correct DM abundance through thermal freeze-out (WIMP ``miracle''). However, conventional WIMPs have become increasingly constrained by indirect/direct detection, and collider experiments, see, e.g.,~\cite{Arcadi:2017kky,Bertone:2018krk}
and references therein. This has led to the exploration of alternative DM candidates beyond the WIMP paradigm. For example, asymmetric dark matter (ADM) \cite{Nussinov:1985xr, Barr:1990ca, Kaplan:1991ah, Kaplan:2009ag,Davoudiasl:2010am,Zurek:2013wia, adm} is an alternative inspired by the DM-baryon coincidence. In this framework, the DM particle is distinct from its antiparticle, and an asymmetry in their respective population densities is generated in the early universe. The core idea of ADM is based on relating and generating DM and baryon/lepton asymmetries through shared interactions. With the exception of a few mechanisms (e.g. \cite{admfromlepto,Buckley:2010ui,Cui:2011qe,Hall:2021zsk}) %
the observed coincidence is achieved with ADM masses of $\mathcal{O}(\text{GeV})$.

Recently, there have been attempts at unifying WIMP DM and ADM mechanisms \cite{wimpyBG, McDonald:2011zza, Davidson:2012fn, wimpyBG2, Cui:2015eba, Farina:2016ndq, Racker:2014uga, Cui:2013bta, Goudelis:2021lra,Cui:2020dly}. Among the existing WIMP-related proposals, \cite{McDonald:2011zza} is highly sensitive to various initial conditions, while both \cite{Davidson:2012fn} and WIMP DM annihilation triggered ``WIMPy baryogensis" \cite{wimpyBG} have sensitivity to ``washout'', i.e., a reduction in the baryon asymmetry due to inverse decays or L-violating scatterings.
The mechanism of ``Baryogenesis from meta-stable WIMPs'' \cite{wimpyBG2} was then proposed as an alternative where the prediction is more robust against model details: the baryon asymmetry is generated by a long-lived WIMP that undergoes CP- and B-violating decays after its thermal freeze-out. Such models also provide a strong cosmological motivation for long-lived particle searches at collider experiments and have become a benchmark for related studies \cite{Cui:2014twa, Cui:2016rqt, ATLAS:2019ems}. However, in its original form, it does not involve the specifics of DM, except for assuming it is WIMP-like. The more recently proposed ``WIMP Cogenesis"\cite{Cui:2020dly} is a new realization of ADM with DM specifics explicitly incorporated, and fully inherits the desirable qualities of both WIMP and ADM production mechanisms. However, the permitted DM mass which produces the observed DM-baryon coincidence is limited to a few GeV. 

From the model building perspective it is desirable to further develop a framework which incorporates the merits of WIMP-like and ADM-like mechanisms. Specifically, a mechanism which maintains a symmetric DM candidate permitting a wider range of DM masses, $\mathcal{O}({\rm GeV-TeV})$, while supplying a predictive connection between $\Omega_{\rm DM}$ and~$\Omega_B$ presents a worthwhile theory target. 
In this work, we consider a realization of this goal in the framework of an isolated dark sector which may have a thermal temperature different from the Standard Model (SM) photon bath (see e.g.,~\cite{Ackerman:mha, Feng:2008mu,Chu:2011be, Chacko:2015noa,Dror:2016rxc,Adshead:2016xxj,Asadi:2021bxp}). In particular, the dark sector is composed of a stable state as DM and a meta-stable dark partner which is the ``parent" of the baryon asymmetry. Both states are initially in equilibrium with each other through the efficient annihilation of DM into dark partners, before freezing out. 
The dark partners are long-lived, and their decays into SM particles trigger the generation of baryon asymmetry.
The stable and meta-stable states, and consequently, $\Omega_{\rm DM}$ and $\Omega_B$, are connected through the overall number-conservation in the dark sector as manifest in the annihilation process. 

Independent of the baryogenesis motivation, we note that the particular dark freeze-out scenario we consider in this work is new to the literature. While there are existing studies on freeze-out in a decoupled dark sector~\cite{Carlson:1992fn, Hochberg:2014dra, Pappadopulo:2016pkp, Dror:2016rxc, Asadi:2021bxp}, some also involving meta-stable annihilation final states, the present scenario distinguishes itself from known possibilities. For instance, in the Co-Decaying dark matter case \cite{Dror:2016rxc}, the meta-stable states decay during the freeze-out process which exponentially depletes the DM abundance, unlike in our case. We provide both numerical and analytical approaches for tracking the dynamical evolution during the freeze-out going beyond the instantaneous freeze-out approximation. 

\begin{figure} 
\centering
\includegraphics[width=0.7\textwidth,trim=0 150 0 150, clip]{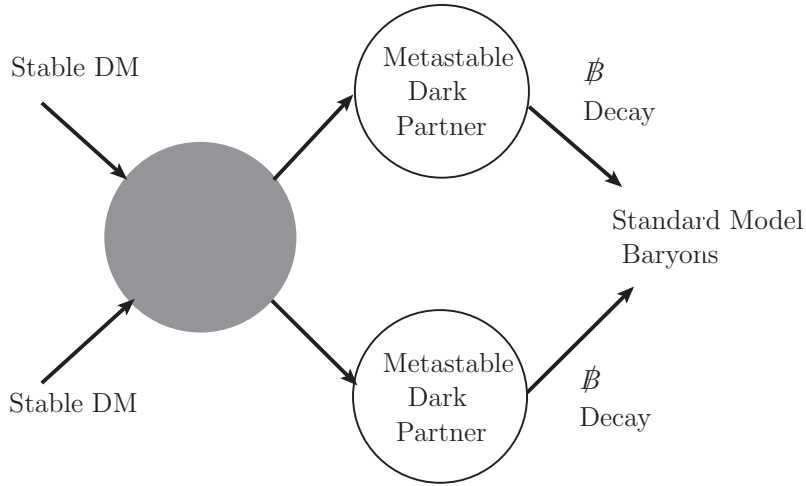}
\caption{Schematic diagram outlining the key stages in the dark freeze-out cogenesis mechanism.}
\label{fig:cartoon}
\end{figure}

The basic idea of \textit{Dark freeze-out Cogenesis} is illustrated in the schematic of Fig. \ref{fig:cartoon}. As we will elaborate later in the paper, a simple realization consists of three dark states $\chi_{1,2,3}$, where $\chi_1$ is taken to be the stable DM candidate and $\chi_2$ is the meta-stable component whose baryon number- and CP-violating decays trigger baryogenesis. Departure from equilibrium is achieved when $\chi_1$ and $\chi_2$ freeze-out in the isolated sector. Finally, $\chi_3$ is required to induce CP-violation.
When the meta-stable $\chi_2$ component %
undergoes CP- and baryon number-violating decays, its abundance is converted into the baryon asymmetry. The %
latter is directly tied to the stable DM component's abundance through number conservation in the dark sector. We demonstrate our general ideas by introducing a UV complete model, i.e., a particle content and their interactions solely based on renormalizable interactions. 

The remainder of the paper is organized as follows: in Sec.~\ref{sec:freeze-out} we outline two scenarios of %
dark sector freeze-out and describe the dynamics of the cosmological evolution of matter abundances around the dark freeze-out. %
In Sec.~\ref{sec:BG}, we discuss the dynamics related to the subsequent $\chi_2$ decays and the ensuing production of the baryon asymmetry. A UV-complete particle physics model is introduced in Sec.~\ref{sec:model}. Various phenomenological implications and observational signatures predicted by the model are detailed in Sec.~\ref{sec:signatures} before concluding in Sec.~\ref{sec:conclusion}. Further details on the freeze-out calculation are relegated to App.~\ref{app:Boltzmann}.

\section{Dark Sector Freeze-out}\label{sec:freeze-out}

In this section, we consider a dark sector with states $\chi_{1}$ and $\chi_2$ that evolve independently from the thermal SM bath. 
This is commensurate with initial conditions where $\chi_{1,2}$ decouple from SM while still being relativistic. 
 The state~$\chi_1$ is stabilized by a ${\mathcal Z}_2$ symmetry, so that it plays the role of the DM candidate. In contrast, a lighter state $\chi_2$ is meta-stable, and decays out-of-equilibrium after $\chi_1$ freezes out, leading to the observed baryon asymmetry (once interference effects with a state~$\chi_3$, to be introduced later, are taken into account). The dark sector is assumed to be self-thermalized before freeze-out, characterised by a common dark temperature $T'$, which is set to be comparable to or lower than the photon temperature $T$, $T'\leq T$, so that the SM sector always dominates the radiation energy of the Universe.
 
Depending on the mass hierarchy between $\chi_1$ and $\chi_2$, there are various options for the details of DM freeze-out. In the following, we focus on two principal possibilities, for each of which we present analytical solutions: $m_{\chi_1} \gg m_{\chi_2}$\,(``hierarchical scenario'') and $m_{\chi_1} = m_{\chi_2} /(1- \delta) $\, with $\delta \ll 1 $ (``nearly degenerate scenario'').\footnote{``Forbidden freeze-out''~\cite{DAgnolo:2015ujb} with $\delta  < 0$ is possible if $\chi_2$ is sufficiently short-lived; we will not discuss this possibility here. 
}
Throughout the paper, we use the subscripts, %
``$i$'', ``f.o.'', and ``$f$''
to represent the initial value, the value at $\chi_1$ freeze-out, and the value well after $\chi_1$ freeze-out, respectively. The subscript ``$0$'' is used for present values (except for $\sigma_0$ below), while the superscript $'$ denotes dark sector quantities that are measured in units of~$T'$.

\subsection[Hierarchical scenario: $ \chi_1$ DM with $m_{\chi_1} \gg m_{\chi_2}$]
{Hierarchical scenario: \boldmath$\chi_1$ DM with \boldmath$m_{\chi_1} \gg m_{\chi_2}$}\label{sec:scenario1}

In the hierarchical scenario, we assume that DM~$\chi_1$ is significantly heavier than $\chi_2$, and reaches its final abundance via non-relativistic freeze-out through $\chi_1 \chi_1 \to \chi_2 \chi_2$ annihilation while $\chi_2$ remains relativistic. 
For the parameters of our interest, this generally requires $m_{\chi_2}\lesssim m_{\chi_1}/20$.
We introduce a time-dependent function, $\xi$, to characterize the temperature ratio between the dark sector ($T'$) and the SM ($T$), with its initial value smaller than unity, $\xi_i \lesssim 1 $. Given that the number of particle species in the SM is much more than in the dark sector we consider, this ensures that the Universe's expansion is solely driven by the visible sector particle content during the radiation-dominated epoch.

While the specific dark freeze-out scenario we consider was not studied in the literature, the analysis of the Boltzmann equation evolution is analogous to some of the existing work on freeze-out in a decoupled dark sector which are useful references for our case, e.g.,~\cite{Ackerman:mha, Feng:2008mu,Chu:2011be, Chacko:2015noa}. In particular, here we use the approach in \cite{Chu:2011be} to calculate the final abundance in terms of the yield variable, $Y_j\equiv n_j/s$, where $n_j$
is the number density of species $i$ and $s = (2\pi^2/45)g_{*S} T^3$ is the {\it total} (dark and visible sector) entropy density, which, by assumption, essentially coincides with the SM one; $g_{*S}$ ($g_*$) are the effective degrees of freedom in entropy (energy) and we take $g_*=g_{*S}$ before neutrino decoupling. 

The thermally averaged annihilation cross section $\langle\sigma_{\chi_1}v\rangle $ is conventionally parametrized as
\begin{align}
  \langle \sigma_{\chi_1} v\rangle \equiv \sigma_0 (x')^{-n} = \sigma_0 x^{-n} \xi^n,
\end{align}
where $x^{(')}= m_{\chi_1}/T^{(')}$ and
 $\sigma_0$ is a reduced ``$n$-wave'' annihilation cross section of $\chi_1$ with $n=0,1,2,\dots$ corresponding to $s,p,d,\dots$-wave annihilation with a respective relative velocity scaling as $v^{2n}$ in $(\sigma_{\chi_1} v)$ before taking the non-relativistic thermal average. Concretely, we obtain
\begin{equation}
Y_{\chi_1, f} ={n+1 \over \lambda } x_\text{f.o.}^{n+1} \xi^{-n}\,,
\label{eq:ychiS1}
\end{equation}
where  $\lambda=\left[s\langle\sigma_{\chi_1}v\rangle /H\right]|_{x = 1}  =0.264 g_{*S}/g_{*}^{1/2}m_{\text{Pl}}m_{\chi_1}\sigma_0$ with $H|_{x=1}=H(m_{\chi_1})$ being the Hubble expansion rate evaluated at a photon temperature $T=m_{\chi_1}$; $m_\text{Pl}=1.2\times10^{19}~\text{GeV}$ is the Planck mass. 
The freeze-out point, $x_\text{f.o.}$, measured in terms of photon temperature, is given by
\begin{eqnarray}
x_\text{f.o.} \equiv {m_{\chi_1} \over T_\text{f.o.}} &=&\xi \ln\left[0.038 \, \xi^{5/2}\langle\sigma_{\chi_1} v\rangle m_{\rm Pl}m_{\chi_1}
\frac{g_{\chi}}{\sqrt{g_*}} c(c+2)\right]\nonumber\\
&&- \frac{\xi}{2} \ln\left\{\xi \ln\left[0.038 \, \xi^{5/2}\langle\sigma_{\chi_1} v\rangle m_{\rm Pl}m_{\chi_1} 
\frac{g_{\chi}}{\sqrt{g_*}}c(c+2)\right]\right\} \,,
\label{eq:xfS1}
\end{eqnarray}
Here, $c= 0.3$ yields good agreement with the exact numerical solution for $\xi \ge 0.01$.
It is worth mentioning that although the temperature ratio has been treated as a constant function above, one should in practice take its value at freeze-out (referred to as $\xi_\text{f.o.}$), which is slightly larger than an initially attained temperature ratio $\xi_i$: $\chi_1$ annihilation heats up the dark sector, and increases this ratio by a factor $2^{1/3}$ as can be obtained from the conservation of entropy in the self-thermalized dark sector. In practice we adopt a step function for the temperature ratio, $\xi$, w.r.t. $x'$ as: $\xi (x') = \xi_i \text{~for }x' \le 4$ and $ 2^{1/3}\xi_i \text{~for }x' \ge 4$. Non-relativistic freeze-out mostly happens at $x' \ge 4$, where most of the entropy density in dark sector is stored in $\chi_2$~\cite{kolbturner}.

Since $x_\text{f.o.} = {m_{\chi_1} /( T'_\text{f.o.}/\xi_\text{f.o.} )} = (2^{1/3}\xi_i) x'_\text{f.o.} $, one may re-write the solution to the DM abundance as
\begin{equation}\label{eq:chi1ab}
 Y_{\chi_1, f} =(2^{1/3}\xi_i)\,\left({n+1 \over \lambda } x_\text{f.o.}^{\prime n+1} \right) \propto {2^{1/3}\xi_i \over \langle\sigma_{\chi_1}v\rangle }\,,
\end{equation}
in analogy to the standard thermal freeze-out solution:  $ Y_{\chi_1, f} = {(n+1) } \, x_\text{f.o.}^{n+1}/ \lambda $. In addition, since the freeze-out point $x'_\text{f.o.}$, measured in units of the dark temperature $T'$ and broadly in the range 3-30, is not sensitive to mild changes of parameters, an $s$-wave cross section $\langle \sigma_{\chi_1} v \rangle /(2^{1/3}\xi_i) \sim {\mathcal{O}}(1)$\,pb is needed to achieve the observed DM abundance in an otherwise standard cosmological thermal history. In case of later co-moving entropy increase by a factor $\Delta$ (to be discussed later), the annihilation cross section should be comparatively smaller, satisfying $\langle \sigma_{\chi_1} v \rangle\, \Delta / (2^{1/3}\xi_i) \sim {\mathcal{O}}(1)$\,pb, in order to generate the observed DM relic density. Compared to the standard case $T' = T$, dark freeze-out with $(2^{1/3}\xi_i) /\Delta < 1$ needs to happen at a relatively higher SM temperature $T$ in order to realize the correct relic abundance. 

For example, considering the interaction term $ig_j \bar \chi_j \gamma^5 \chi_j A $ ($j =1,\,2$) induced by a heavy pseudoscalar state~$A$ (this is one of the interactions realized in our UV complete model presented in Sec.~\ref{sec:model}), the annihilation cross section is dominantly $s$-wave and given by 
\begin{equation}\label{eq:psAnni}
  \sigma_{\chi_1} v = {g_1^2 g_2^2 \sqrt{s (s-4m_{\chi_2}^2)} \over 16\pi (s-m_A^2)^2} \to {g_1^2 g_2^2  m_{\chi_1}^2 \over 4\pi m_A^4} \,,
\end{equation}
where the limit relates to non-relativistic relative initial motion with $m_A\gg 2m_{\chi_1}$. Therefore, in order to generate the observed DM abundance, this points to 
\begin{equation}
 m_{\chi_1} \sim 0.18\,\text{TeV} \, \left({ 0.1 \over g_1 g_2 }\right)\, \left({m_A \over {\rm TeV} }\right)^2 \,\left({ \xi_\text{f.o.}/\Delta \over 0.01}\right)^{1/2}\,. 
\end{equation}

Since $\chi_2$ particles remain relativistic during the freeze-out (in the case of $m_{\chi_2}/m_{\chi_1}\lesssim x'_\text{f.o.}$), their abundance is fixed in terms of the temperature ratio as $Y_{\chi_2, f}=0.42 g_{\chi}\xi_i^3/ g_{*} - Y_{\chi_1, f}$. This is because in our setup, by assumption, all interactions in the dark sector preserve the {\it total} number of $\chi_1$ and $\chi_2$, such that $Y_{\chi_1}+Y_{\chi_2} = \text{const.}$, up to changes in co-moving entropy.

In the end, the observed baryon asymmetry $Y_B$ produced from $\chi_2$ decays is proportional to the ``would-be'' $\chi_2$ abundance $Y_{\chi_2, f}$ if it had not decayed: $Y_{B} \propto \epsilon_{\rm CP} Y_{\chi_2, f}$, where $\epsilon_{\rm CP}$ is the CP asymmetry of the decay and $\tau$ is the $\chi_2$-lifetime at rest. A later entropy increase, which may be induced by the late decay of $\chi_2$, would dilute the $\chi_2$ abundance, requiring a larger $\epsilon_{\rm CP}$ to yield the observed baryon asymmetry today. More explanation about the final yield of matter abundances will be given in Sec.~\ref{sec:model}.

\subsection[Nearly degenerate scenario: $\chi_1$ DM with $m_{\chi_1} = m_{\chi_2} + \delta m $]{Nearly degenerate scenario: \boldmath$\chi_1$ DM with \boldmath$m_{\chi_1} = m_{\chi_2} + \delta m $}\label{sec:scenario2}
 
In the second, nearly degenerate scenario, we assume that $\chi_1$ is marginally heavier than $\chi_2$ and $\chi_1$ is still stabilized by a ${\mathcal Z}_2$ symmetry. The mass degeneracy is characterised by a dimensionless parameter, $\delta \equiv (m_{\chi_1}- m_{\chi_2})/m_{\chi_1} \ll 1$. The final abundances of DM candidate $\chi_1$ and of meta-stable $\chi_2$ are dominantly set by non-relativistic freeze-out of $\chi_1\chi_1\leftrightarrow\chi_2\chi_2$. 
The crucial difference from the previous scenario is that $\chi_2$ is now non-relativistic during the freeze-out of~$\chi_1$. This, in turn, leads to a faster decrease of dark temperature. 
While dark freeze-out scenarios involving two near-degenerate states have been discussed in literature
(see e.g.,~\cite{Dror:2016rxc, Okawa:2016wrr, Kopp:2016yji, Bernal:2017mqb, DAgnolo:2018wcn, Maity:2019hre}), in our scenario the decay of the meta-stable state occurs well after the freeze-out era, resulting in different dynamics. Here we provide an analytical solution to this problem, which to our knowledge is novel.

 Due to the mass degeneracy, the difference from the earlier discussed hierarchical scenario is that, both $\chi_1$ and $\chi_2$ are non-relativistic at DM freeze-out. At the same time, the total co-moving number of $\chi_1$ and $\chi_2$ particles is still conserved, and the ratio of $n_{\chi_1}$ and $n_{\chi_2}$ in equilibrium depends on their mass splitting and the dark temperature. 
In this scenario, the temperature ratio $T'/T$ can deviate significantly from its initial value during Universe expansion, so we must follow both number and energy density in the dark sector. They are described by the respective Boltzmann equations,
\begin{eqnarray} \label{eq:numS2}
\dot{n}_{\chi_1}+3Hn_{\chi_1} = -( \dot{n}_{\chi_2} + 3Hn_{\chi_2} ) =-\langle\sigma_{\chi_1}v\rangle n_{\chi_1}^2+\langle\sigma_{\chi_2}v\rangle n_{\chi_2}^2 \,,
\end{eqnarray}
as well as 
\begin{equation} 
\dot{\rho'}+ 3 H ({\rho'}+p') =0\,.
\end{equation}
where the dot signifies the derivative with respect to cosmic time and $\rho'$ ($p'$) is the energy density (pressure) of the dark sector. For a dark sector in kinetic equilibrium, as assumed here, we are to solve the set of three Boltzmann equations to obtain the dark temperature, $T'$, and two number densities, $n_{\chi_{1,2}}$ (or equivalently, two chemical potentials $\mu_{\chi_{1,2}}$). Note, however, that $\langle\sigma_{\chi_1}v\rangle $ and $\langle\sigma_{\chi_2}v\rangle$ are related through the principle of detailed balance; see~Eq.~\eqref{eq:chi2ann} in the appendix. 

\begin{figure}%
\centering
\includegraphics[width=0.49\textwidth]{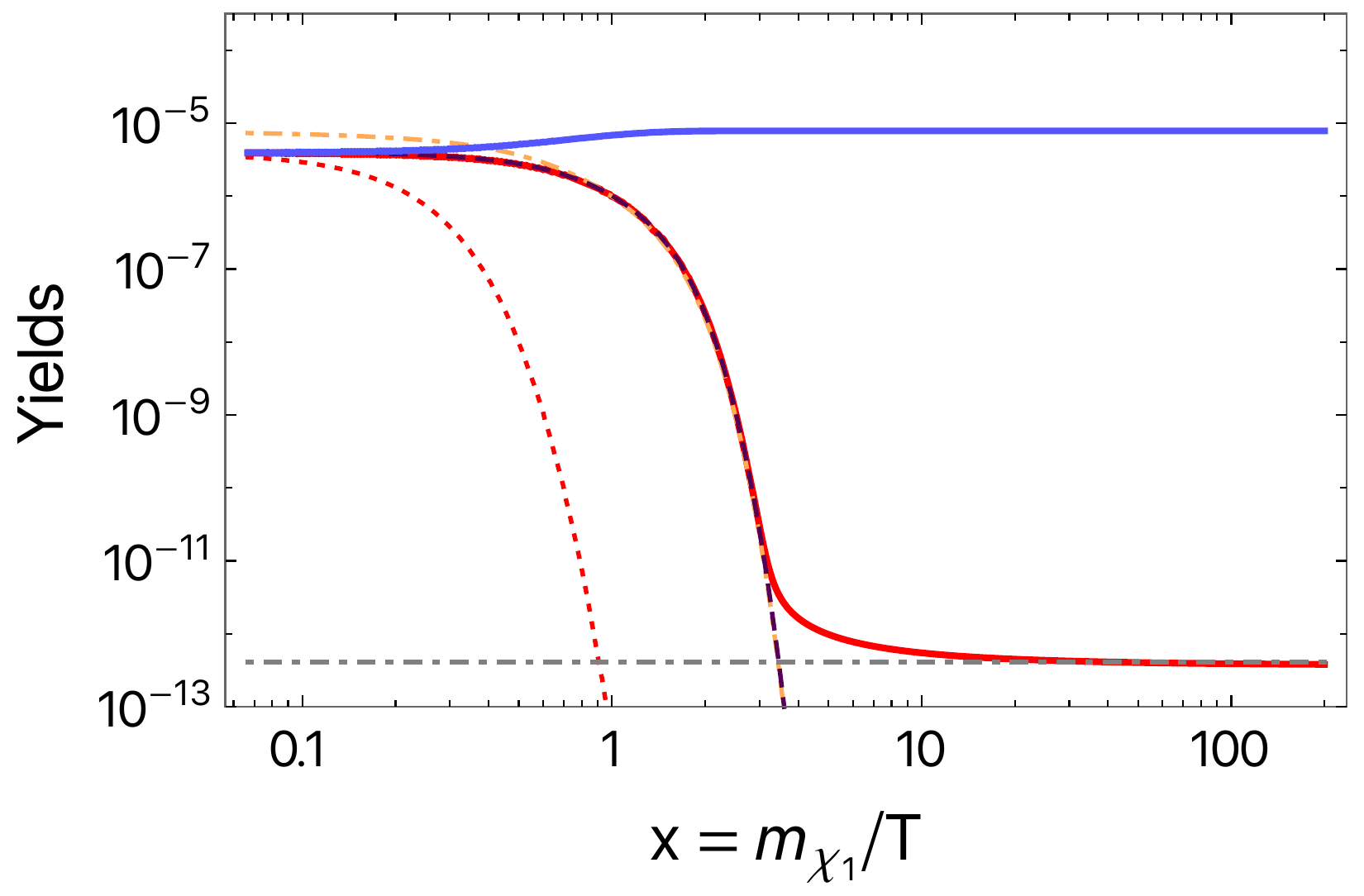}
\includegraphics[width=0.49\textwidth]{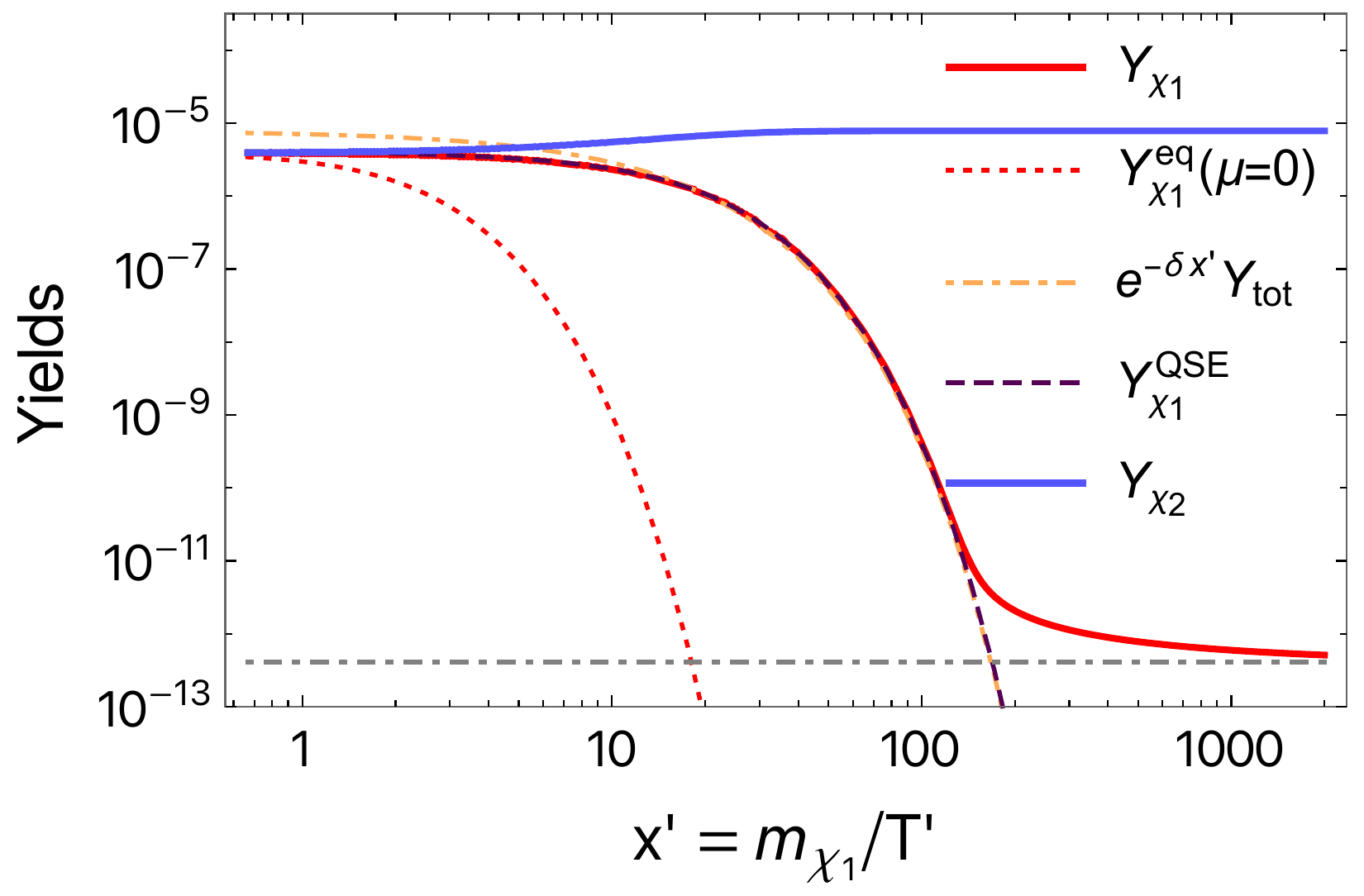}
\caption{The abundance evolution for the near-degenerate case with $m_{\chi_1} = 1$\,TeV and $\delta =0.1$ as a function of $x$ (left panel) and $x'$ (right panel); we assume $\xi_i =0.1$, and $g_1 g_2/m_A^2 =1/(4\,\text{TeV})^2$. The red (blue) solid lines show the  evolution of $Y_{\chi_1}$ ($Y_{\chi_2}$), and the dotted  red lines show the equilibrium value of $Y_{\chi_1}$ calculated from~$T'$ with zero chemical potential. The orange dot-dashed lines show $e^{-\delta x'}Y_{\rm tot} = e^{-(m_{\chi_1}-m_{\chi_2})/T'}(Y_{\chi_1, i}+ Y_{\chi_2, i})$, which is an approximation of the exact quasi-static equilibrium solution, $Y^{\rm QSE}_{\chi_1}$ (purple dashed lines); see Eq.~\eqref{eq:QSE} for more details. Both show that the suppression of $Y_{\chi_1}$ is solely due to the mass splitting, independent of the absolute mass of ${\chi_1}$.  The observed DM abundance is shown as gray dot-dashed lines.} 
\label{Fig:Split01}
\end{figure}

For illustration, we again take the pseudoscalar mediated model with unaltered annihilation cross section given by Eq.~\eqref{eq:psAnni}, and solve the set of equations above for several benchmark parameters. The results are given in Fig.~\ref{Fig:Split01}. The right panel shows that freeze-out is associated with sizable $x'$ since the Boltzmann suppression of $Y_{\chi_1}$ with respect to $Y_{\chi_2}$ is determined by the mass splitting (orange dot-dashed lines),\footnote{The function $e^{-\delta x'}Y_{\rm tot}$ is an approximation to the quasi-static equilibrium solution $Y^{\rm QSE}_{\chi_1}$, defined by Eq.~\eqref{eq:QSE} in the appendix,  in the limit $x'\gtrsim 1/\delta$; the detailed derivation is provided there.} instead of the DM mass (red dotted lines), when both DM annihilation and its inverse process are in equilibrium. This is dictated through overall dark number conservation, and by the fact that the inverse process, $\chi_2\chi_2\to \chi_1\chi_1$, only becomes suppressed once $T' \ll \delta m$. Nevertheless, when the same solutions are shown as a function of $x$ (left panel of Fig.~\ref{Fig:Split01}) one observes that freeze-out happens with $x_\text{f.o.} \sim 3$, smaller than the typical value, $x_\text{f.o.} \sim 20$, in the standard case ($\xi \equiv 1$) of TeV DM freeze-out, to get the observed DM abundance; the reason for it is explained below Eq.~\eqref{eq:chi1ab}. %

While these equations can only be solved numerically, which is how we obtained our final results, here we also provide analytical explanation/approximation to better demonstrate the underlying physics. An empirical relation between $T'$ and $T$ can be obtained using the observation that a decoupled non-relativistic species cools down adiabatically with the scaling $T' \propto T^2$; for a relativistic species $T' \propto T$ instead (away from epochs of entropy transfer). The transition between these two scaling laws happens approximately at $T' \sim m_{\chi_2}/4 $. Using the insights we gained from our numerical study, a good approximation at $T'\ll m_{\chi_2}$ is 
\begin{equation}\label{eq:empicialT}
 { \xi(T) \over 2^{1/3}\xi_i} \approx {2^{1/3}\xi_i T \over m_{\chi_2}/ 4 } = {4 \times 2^{1/3} \xi_i \over x(1-\delta) } \, 
\end{equation}
at any temperature $T$, where $\xi_i$ is the initial temperature ratio, $2^{1/3}$ is the reheating factor discussed earlier. That is, we obtain the relation $x'= \beta x^2$ at $T'\ll m_{\chi_2}$, with the dimensionless constant  $\beta \equiv (1-\delta)/(4\times 2^{2/3}\xi_i^2)$. See Fig.~\ref{Fig:DarkTapp} in the Appendix for a comparison with full numerical results.

\begin{figure}%
\centering
\includegraphics[width=0.49\textwidth]{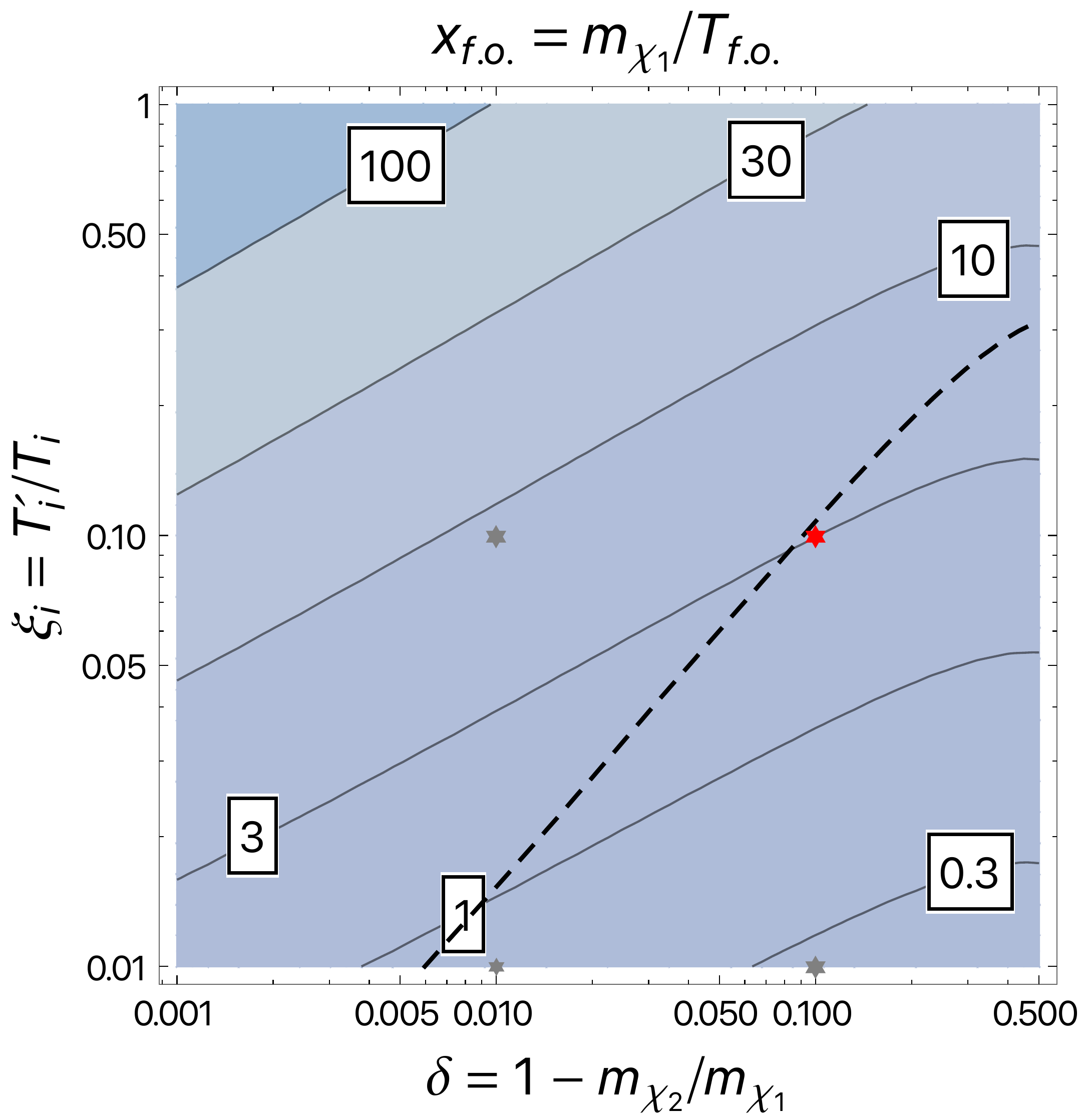}~~
\includegraphics[width=0.49\textwidth]{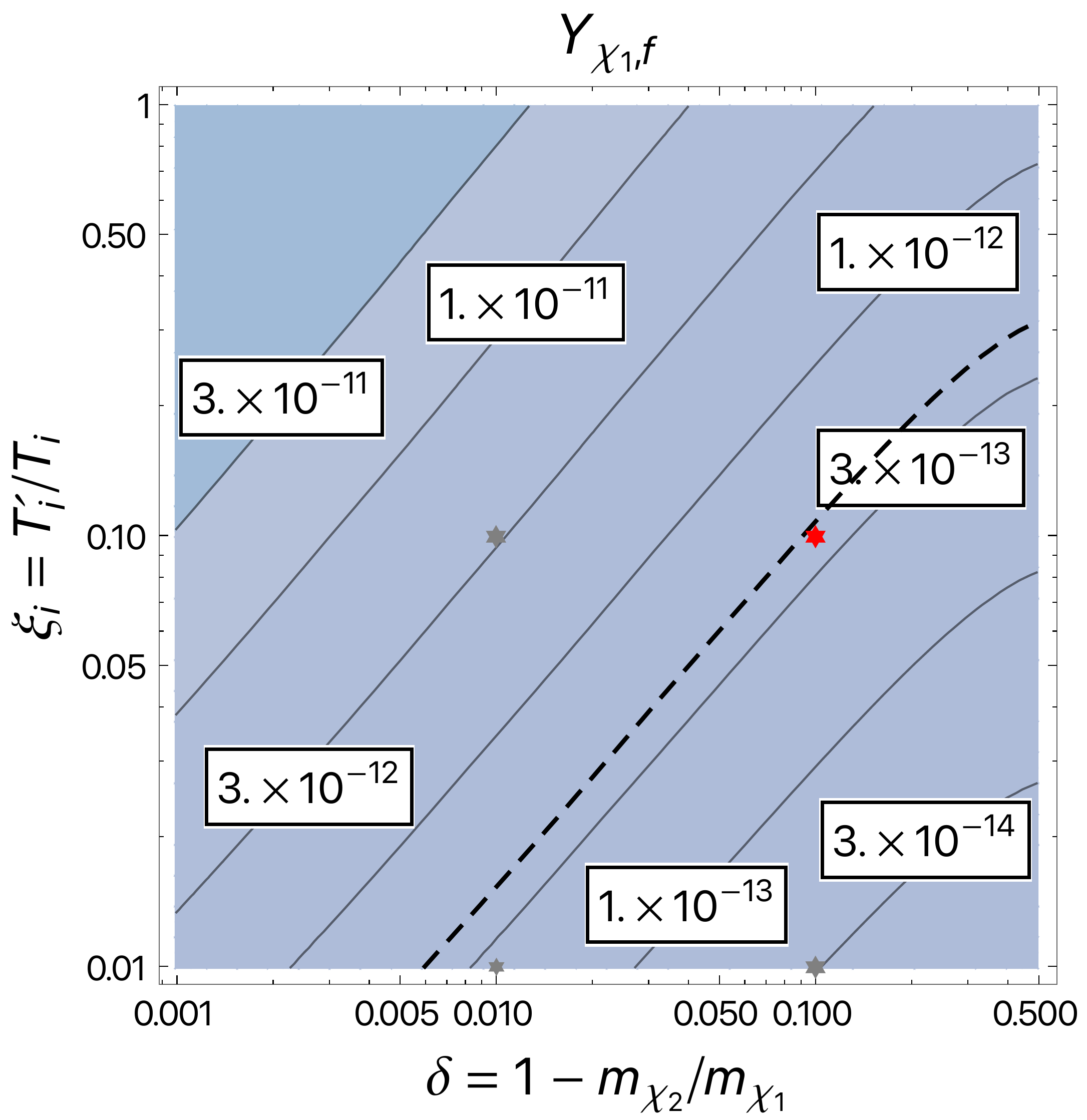}
\caption{Contours of $x_\text{f.o.}$ (left panel) and $Y_{\chi_1,\,f}$ (right panel) obtained from the analytical solution of Sec.~\ref{sec:scenario2} as a function of mass splitting, $\delta$, and the initial temperature ratio between the dark and SM sectors, $\xi_i$. In both panels, we choose $m_{\chi_1} = 1\,$TeV, as well as $g_1 g_2/m_A^2 =1/(4\,\text{TeV})^2$.
The dashed black lines show the observed DM abundance. The red (gray) stars depict the parameter choices used in Fig.~\ref{Fig:Split01} (Fig.~\ref{Fig:SplitAdd}), where the Boltzmann equations are solved numerically for $\delta =0.1$ and $\xi_i=0.1$.}
\label{Fig:anaFO}
\end{figure}

Adopting Eq.~\eqref{eq:empicialT} allows us to solve Eq.~\eqref{eq:numS2} analytically as detailed in App.~\ref{app:Boltzmann}. Following the notation of Eq.~\eqref{eq:ychiS1}, we obtain an approximate solution of a non-relativistic freeze-out,
\begin{equation}\label{eq:chiabund}
	Y_{\chi_1, f} \approx \frac{(2n+1)\beta^n}{ \lambda }x_\text{f.o.}^{2n+1} \,,
\end{equation}
where the freeze-out time is given by
\begin{equation}\label{eq:fotemp}
 x_\text{f.o.}^2 = {1\over \beta \delta } \ln \left[ \frac{ 0.21 c(2+c) \lambda \xi^3_i \beta^{1/2} \delta^{1/2} }{(1-\delta)^{3/2} } \, {g_{\chi} \over g_{*}} \right] - { n+{3\over 2} \over  \beta \delta } \ln \left\{\ln \left[ \frac{ 0.21 c(2+c) \lambda \xi^3_i \beta^{1/2} \delta^{1/2} }{(1-\delta)^{3/2} } \, {g_{\chi} \over g_{*}} \right] \right\}\,.
\end{equation}
The solution is different from Eq.~\eqref{eq:ychiS1} of the previous scenario, mostly due to $T' \propto T^2$ at freeze-out here. Nevertheless, in the $s$-wave limit, we have 
\begin{equation}
  Y_{\chi_1, f} \approx \frac{1}{ \lambda }x_\text{f.o.} \propto {1\over m_{\chi_1} \sigma_0} x_\text{f.o.} \,,
\end{equation} 
in both scenarios. It in turn suggests that $\Omega_{\rm DM}$ is generally not sensitive to the DM mass, as it cancels in the ratio $x_\text{f.o.}/m_{\chi_1}$ and for as long as the $s$-wave annihilation cross section $\sigma_0$ is held constant. Hence, the yield decreases for increasing $\sigma_0$.

The analytical solutions are illustrated in Fig.~\ref{Fig:anaFO} by scanning the parameter space for $m_{\chi_1} = 1$\,TeV and $g_1 g_2/m_A^2 =1/(4\,\text{TeV})^2$. For $\xi=0.1$ and $\delta=0.1$, it gives the freeze-out point $x_\text{f.o.} \approx 3$, as well as the observed DM abundance, in good agreement with the exact numerical results of Fig.~\ref{Fig:Split01}. For additional numerical results of $n_{\chi_{1,2}}$ evolution with various parameter sets, see Fig.~\ref{Fig:SplitAdd} in the appendix. For instance, reducing the mass splitting by a factor of $10$ to $\delta=0.01$ gives $x_\text{f.o.} \approx 8$ and a final abundance about ten times larger (middle gray star in Fig.~\ref{Fig:anaFO}). This can be explained as follows: estimating the freeze-out condition through $(Y_{\chi_1}s)\langle \sigma_{\chi_1}v \rangle\sim H$, one obtains $Y_{\chi_1}\sim H/(s \langle \sigma_{\chi_1}v \rangle) \propto 1/T$ at freeze-out. That is, smaller mass splitting leads to a lower freeze-out temperature, and thus larger value of the final $ {\chi_1}$ yield.
In contrast, reducing the temperature ratio gives an less populated dark sector, and thus a smaller final abundance of $\chi_1$. 

Finally, we turn to the $\chi_2$ abundance. Number conservation of the sum of $\chi_1$ and $\chi_2$ directly translates into $Y_{\chi_2, f} \approx 0.42\xi^3_i g_{\chi}/g_{*} - Y_{\chi_1, f}$ at the end of freeze-out. Eventually, we demand $\chi_2$ to decay into the visible sector, in analogy to the hierarchical scenario before.

\subsection[Potential matter-domination caused by $\chi_2$]{Potential matter-domination caused by \boldmath$\chi_2$}
\label{sec:EMD}

In both the hierarchical and near-degenerate scenarios described above, at the end of DM freeze-out the $\chi_2$ abundance is given by 
\begin{equation}
 Y_{\chi_2, f} \approx {0.42 g_{\chi}\xi_i^3 \over g_{*}} - Y_{\chi_1, f} \,. 
\end{equation}
As we focus on $\xi_i \ge 10^{-2}$ and the DM abundance should satisfy $Y_{\chi_1, f} \approx 10^{-10}(4\,\text{GeV}/m_\phi) < 10^{-10} $ for $m_{\chi_1}$ well above GeV, the equation above gives  $Y_{\chi_2, f} \gg Y_{\chi_1, f}$, thus the second term on the R.H.S. can be neglected in practice. Note that the insensitivity of $Y_{\chi_2, f}$ to the annihilation cross section is a result of the dark number conservation and the Boltzmann suppression of $Y_{\chi_1, f}$ relative to $Y_{\chi_2, f}$ around freeze-out time due to the mass splitting.
Interestingly, depending on the lifetime of $\chi_2$, the $\chi_2$ abundance, now mainly decided by $\xi_i$, may lead to an epoch of early matter-domination (EMD). This would modify the prediction for current-day particle abundances, due to entropy injection at the end of EMD. 

\begin{figure}%
\centering
\includegraphics[width=0.49\textwidth]{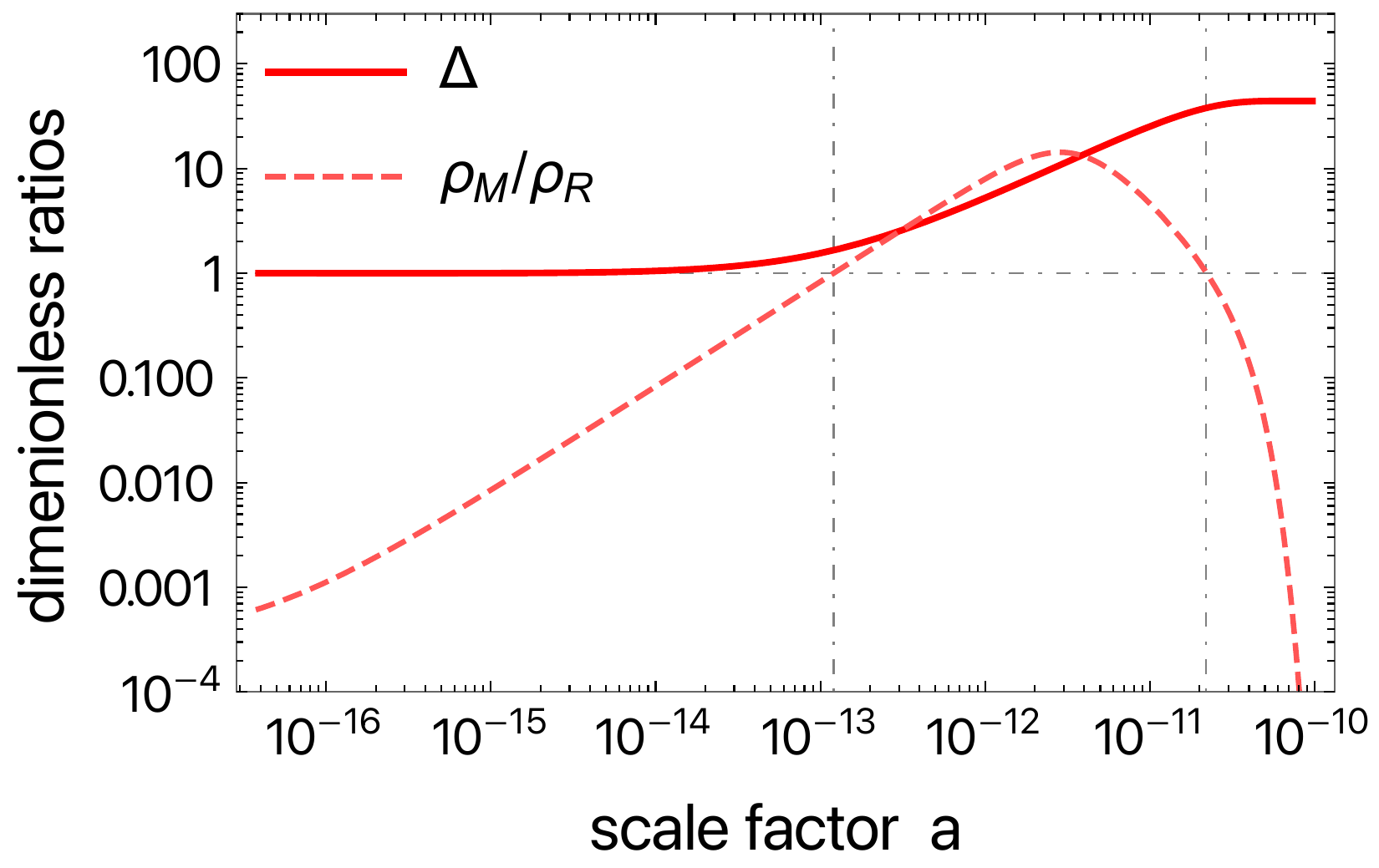}
\includegraphics[width=0.49\textwidth]{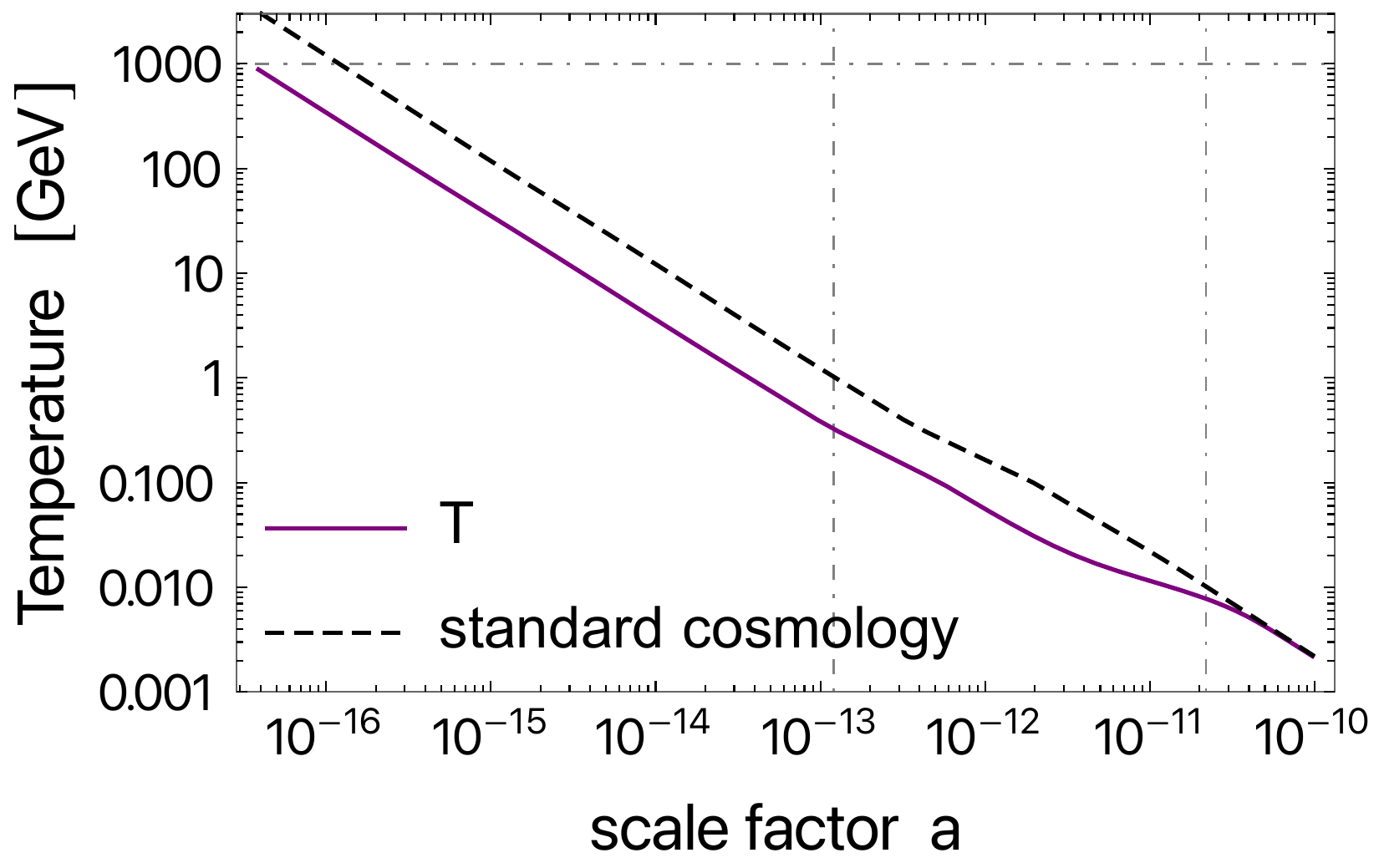}
\caption{Evolution of the dark sector with an early matter-domination epoch as a function of the scale factor (normalized to a present value $a_0=1$) for $m_{\chi_2} = 1\,$TeV, $\tau =10^{-2}$\,sec, and $\xi_i = 0.3$. The left panel shows the evolving ratio of matter-to-radiation energy densities, $\rho_{\rm M}/\rho_{\rm R}$ (dashed line), and $\Delta$--the ratio of entropy density defined by $a^3(\rho_{\rm M}+ \rho_{\rm R})^{3/4}$ to its initial value (solid line); see Eq.~\eqref{eq:defineDelta}. In the right panel, the purple solid (black dashed) line shows the evolution of the SM temperature $T$ in our model (in standard cosmology). In both panels, the vertical gray lines depict the beginning (end) of this epoch at $a_\alpha$ ($a_\beta$), while horizontal gray lines correspond to unity in left panel, and  $T = m_{\chi_2}$ in right panel.}
\label{Fig:SimpleEvo}
\end{figure}

If $\chi_2$ decays while still being  relativistic, the Universe's expansion, by assumption, is dominated by SM radiation, and the final baryon asymmetry remains unaltered from Eq.~\eqref{eq:Bab}. 
In contrast, the decay of a non-relativistic species transfers additional entropy to the SM sector. The dilution factor, measured in terms of the ratio of total co-moving entropies before ($S_\alpha$) and after ($S_\beta$) decay
may be estimated as
\begin{equation}\label{eq:defineDelta}
	\Delta \equiv  { S_\beta \over  S_\alpha} =  {a_\beta^3 \over a_\alpha^3} { s_\beta \over s_\alpha} = {a_\beta^3 \over a_\alpha^3} \left({\rho_\beta \over \rho_\alpha}\right)^{3/4}\,,
\end{equation}
where the subscripts, $\alpha$ and $\beta$, mark the beginning and end points when the Universe exits and re-enters the radiation-domination, respectively. Note that the radiative entropy ($S$) does not include the non-relativistic matter contribution; the last quantity is thereby used during the EMD epoch. 
Using the energy conservation law $\dot \rho = -3H (\rho+ p)$ and the equation of state $p = w\, \rho$, one obtains
\begin{equation}
  {\rho_\beta \over \rho_\alpha} = \exp\left(-3\int^{a_\beta}_{a_\alpha} da \, {1+\omega \over a }\right) \,,
\end{equation}
where $\omega = 1/3$ ($0$) for a radiation-dominated (matter-dominated) Universe. During the transition, $\omega$ varies and needs to be calculated numerically. The results for one benchmark parameter set is shown in Fig.~\ref{Fig:SimpleEvo}.

We may also obtain a simplified analytic estimate for the dilution factor $\Delta$ by assuming a sudden transition between matter- and radiation-dominated epochs. For this, one may take $\rho \propto a^{-3}$ during matter domination,  thus $\Delta \approx (a_\beta/a_\alpha)^{3/4}$ from Eq.~\eqref{eq:defineDelta}, i.e., $\Delta$ is mostly decided by the duration of this period. 
Matter-domination starts at the scale factor $a = a_\alpha$, when $ (Y_{\chi_2, f} s ) m_{\chi_2} \approx \rho_{\rm SM}$ with a photon temperature $T(a_\alpha) \approx 4Y_{\chi_2, f} m_{\chi_2}/3$, as well as a matter (or radiation) energy density $\rho_M(a_\alpha) \approx g_* (Y_{\chi_2, f} m_{\chi_2})^4$ at this time. The matter-domination epoch ends when the Hubble rate equals the $\chi_2$ decay rate, and $\rho_M(a_\beta) = 3\Gamma_{\chi_2}^2/(8\pi G)$ using $\Gamma_{\chi_2} = H(a_\beta)$. 
Therefore, we obtain 
 \begin{equation}
 	\Delta \approx \max\left\{ 1, \, \left({\rho_M(a_\alpha) \over \rho_M(a_\beta)}\right)^{1/4}\right\} \approx \max\left\{ 1, \, 0.7 \xi_i^3 \,{g_{\chi_2} \over g_*^{3/4} } \left({m_{\chi_2} \over \sqrt{m_{\rm pl}\ \Gamma_{\chi_2}}}\right)\right\} \,. 
 	\label{Eq:dilution}
 \end{equation}
Evidently, when $\Delta \approx 1$, the matter-domination epoch was either very brief or never happened. The expression above can be tested against the numerical solution of Fig.~\ref{Fig:SimpleEvo}.\footnote{Note that the decoupling of heavy SM particles contributes to the evolution of $T$ in the right panel of Fig.~\ref{Fig:SimpleEvo}, but it does not affect the entropy evolution in the left panel. } For example, while Eq.~\eqref{Eq:dilution} yields $\Delta \approx 41$, Fig.~\ref{Fig:SimpleEvo} points to $\Delta \approx 43$. For better guidance, we point out that $\Delta \lesssim 10^3$ for $m_{\chi_2}$ smaller than 1\,TeV, due to $g_{\chi_2} \ll g_*$ and the Big Bang Nucleosynthesis (BBN) bound on the lifetime of ${\chi_2}$.
 
In our setup, the EMD epoch occurs after DM freeze-out. Therefore, while the dilution reduces the values of $Y_{\chi_1}$ and $Y_{\chi_2}$ separately by a common factor $\Delta$, 
it does not change their ratio. That is, in absence of other baryon asymmetry-depleting processes we have,
\begin{equation}\label{eq:Bab}
 Y_{\chi_1, 0 }= { Y_{\chi_1, f} \over \Delta} \text{~~~ and ~~~~} Y_{B }= \epsilon_{\rm CP} { Y_{\chi_2, f} \over \Delta} \,,
\end{equation}
and the predicted relation between the DM abundance and baryon asymmetry is not affected.

\section{\boldmath$\chi_2$ Decays, Baryogenesis, and the DM-Baryon Coincidence}\label{sec:BG}
As outlined earlier, the $\chi_2$ abundance following the dark freeze-out is meta-stable, and its CP- and B- violating, out-of-equilibrium decays trigger baryogenesis. We require that $\chi_2$ decays after freeze-out of both $\chi_1$ and $\chi_2$, but before primordial nucleosynthesis: $T_{\rm BBN}<T_{\chi_2,\text{dec}}<T_\text{f.o.}$. This way, freeze-out and baryogenesis can be treated separately, and the wash-out effect from inverse decay of $\chi_2$ is ineffective during the baryogenesis stage. 
 Washout processes such as $udd\rightarrow\bar{u}\bar{d}\bar{d}$ are suppressed by the large mediator mass (required by collider constraints, see Secs.~\ref{sec:model} and \ref{sec:signatures}) in the effective operator for the process. A more detailed discussion on suppressing various washout processes can be found in \cite{wimpyBG2} which is in analogy to the scenario considered here. %
 
The initial condition for $Y_{\chi_2}$ of this stage of evolution (i.e., $\chi_2$ decay) is set by the would-be abundance of $\chi_2$ after the freeze-out of $\chi_1$, which is an essential factor for predicting the baryon asymmetry. Recall from Sec.~\ref{sec:EMD}, via the overall number conservation within the dark sector, the $\chi_2$ abundance after the freeze-out of $\chi_1\chi_1\leftrightarrow\chi_2\chi_2$ is given by $Y_{\chi_2, f}=(Y_{\chi_1,i}+Y_{\chi_2,i})-Y_{\chi_1,f}$ where $Y_{\chi_1,i}\approx Y_{\chi_2,i}=0.21g_\chi/g_{*}\xi_i^3$. 
Assuming that it violates baryon number by one unit (which is realized in the model shown in the next section), the co-moving density of baryon asymmetry $Y_B \equiv (n_b - n_{\overline b})/s$ observed today is
\begin{equation}\label{eq:chi2dec}
Y_{B}=\epsilon_{CP}\int_{0}^{T_\text{f.o.}}\frac{dY_{\chi_2}}{dT}\exp\left(-\int_{0}^{T}\frac{\Gamma_{W}(\tilde{T})}{H(\tilde{T})}\frac{d\tilde{T}}{\tilde{T}}\right)dT+ \dots\,,
\end{equation}
where $\epsilon_{CP}$ is the CP asymmetry in $\chi_2$ decays, and $\Gamma_{W}$ is the rate of washout processes. The ellipses stand for any possible additional sources of the baryon asymmetry prior to $\chi_2$-decay,\footnote{
An initial contribution $Y_{B}^\text{initial}$ to the baryon asymmetry can be accounted for by the addition of a term $Y_{B}^\text{initial}\exp\left(-\int_{0}^{T_\text{initial}}\frac{\Gamma_{W}(T)}{H(T)}\frac{dT}{T}\right) $ where the exponential suppression is due to ``wash-out.''
} which we simply assume to be zero in the following.

As mentioned earlier, the rate of washout processes that can potentially reduce the baryon asymmetry is weak by construction and the exponential factors can be dropped. This gives a simple approximate solution for the co-moving baryon asymmetry in either, hierarchical or near-degenerate, scenario,
\begin{equation}\label{eq:basym}
Y_{B}\approx \epsilon_{CP}{ Y_{\chi_2, f} \over \Delta} ={ \epsilon_{CP} \over \Delta} \left(0.42\frac{g_{\chi}}{g_{*}}\xi_i^3-Y_{\chi_1, f}\right) ,
\end{equation}
where $Y_{\chi_1, f}$ is given in Eqs.~\eqref{eq:chi1ab} and \eqref{eq:chiabund}. The result in Eq.~\eqref{eq:basym} directly connects the co-moving baryon asymmetry to the relic abundance of $\chi_1$. Assuming that $\chi_1$ composes all of DM observed today, we obtain the ratio of the DM and baryon abundances observed today as  %
\begin{equation}\label{eq:coincidence}
\frac{\Omega_{B}}{\Omega_{\chi_1}}=\epsilon_{CP}\frac{m_p}{m_{\chi_1}}\frac{Y_{\chi_2, f}}{Y_{\chi_1, f}}=\epsilon_{CP}\frac{m_p}{m_{\chi_1}}
\times\begin{cases} 
0.42\frac{g_{\chi}}{g_{*}}\frac{ \xi_i^{3} \lambda}{(n+1)\xi_{\rm f.o.}^{-n}}\frac{1}{x_\text{f.o.}^{n+1}}-1
& \text{hierarchical}\\
0.42\frac{g_{\chi}}{g_{*}}\frac{ \xi_i^{3} \lambda}{(2n+1)\beta^n }\frac{1}{x_\text{f.o.}^{2n+1}}-1
& \text{near-degenerate}
\end{cases}\,,
\end{equation}
where $m_p$ is the proton mass; the degrees of freedom are to be evaluated at the time of DM freeze-out $g_{*} \approx g_{*S}\sim 100$.\footnote{In the numerical evaluation we carry along the temperature dependence of $g_{*}$ and $g_{*S}$, whereas in the analytical estimates we take them as constants.} 
In the case of $s$-wave annihilation ($n=0$), neglecting the second term allows to further simplify the expression to
\begin{equation}\label{eq:swave}
\frac{\Omega_{B}}{\Omega_{\chi_1}} \approx \frac{\epsilon_{CP}}{10^{-9}} \left[ {\xi_i^3  \over x_\text{f.o.}} \frac{g_\chi}{g_{*}^{1/2} } \left( \frac{\sigma_0}{\rm 0.3\, pb} \right)\right]\,.
\end{equation}
The result in Eq.~\eqref{eq:coincidence} for $\Omega_{B}$ takes a similar form as the conventional WIMP miracle, augmented by the CP-asymmetry factor $\epsilon_{CP}$ and proton-to-$\chi_1$ mass ratio $m_p/m_{\chi_1}$. Taking similar values for the annihilation cross section, splitting $\delta$, and dark sector temperature $\xi_i$ to those in Fig.~\ref{Fig:Split01} and Fig.~\ref{Fig:SplitAdd}, the observed coincidence between dark and baryonic matter abundances can be achieved with $m_{\chi_1}\sim\mathcal{O}(10-10^4)~\text{GeV}$ and $\epsilon_{CP}\approx\mathcal{O}(10^{-8}-10^{-2})$.
In the next section, we implement the general dynamics of $\chi_1$-$\chi_2$ freeze-out and $\chi_2$ decay-triggered baryogenesis in a specific model which achieves exactly this with dedicated numerical analysis.

\section{Exemplary UV-complete Model}\label{sec:model}

In this section we present a concrete model with fermionic dark states, which include interactions realizing the $\chi_1\chi_1\rightarrow \chi_2\chi_2$ freeze-out and generation of the baryon asymmetry. For the latter, the three Sakharov conditions~\cite{Sakharov} are met by the CP- and baryon number-violating out-of-equilibrium decay of $\chi_2$ to SM quarks.

We start by considering dark sector interactions necessary for freeze-out. There are a number of options to achieve this and our core mechanism is insensitive to the detailed realization of it. Here, we consider interactions mediated by a pseudoscalar, $A$, or a scalar, $S$. The associated Lagrangian then reads,
\begin{align}
 \mathcal{L}_\text{dark f.o.} & = - g_j\bar{\chi}_{j}i\gamma^5\chi_{j} A - g'_j\bar{\chi}_{j} \chi_{j} S \,, \label{eq:fo}
\end{align}
where the SM singlet Majorana fermions $\chi_j$ ($j=1, 2, 3$) are the members of the dark sector. The interactions mediated by~$A$ allow for $s$-wave annihilation, whereas the leading contribution in a velocity expansion of the annihilation cross section mediated by~$S$ is $p$-wave. 
In order to avoid additional complications of a dynamical role that $\chi_3$ may play during dark freeze-out, we may take   $g^{(')}_3 \ll 1$. This allows us to neglect $\chi_{1,2}\chi_{1,2}\leftrightarrow \chi_3\chi_3$ processes and $\chi_3$-induced energy transfer between dark and SM sectors even for $\beta = O(1)$ ($\beta$ is to be introduced in Eq.~\ref{eq:bg}). 
In addition, for $m_{A(S)}\geq 2m_{\chi_1}$ we may also neglect the processes $\chi_{1,2}\chi_{1,2}\leftrightarrow AA (SS)$.
Taken together, we then map onto the analysis of dark sector freeze-out of the preceding section. 
In passing, we note that a hierarchical coupling structure among $g_j^{(')}$ can be addressed by an underlying dark flavor physics model, but which is beyond the scope of this work.

In order to predict the yields of DM and baryon asymmetry from this model, we first analyze the freeze-out process due to the annihilation $\chi_1\chi_1\leftrightarrow\chi_2\chi_2$ mediated by either the pseudoscalar~$A$ ($s$-wave) or the scalar~$S$ ($p$-wave). The non-relativistic thermally averaged annihilation cross-section (away from the resonances) reads
\begin{align}\label{eq:sannihilation}
 \sigma_{\chi_1}v \approx\frac{g_1^2g_2^2\sqrt{\delta(2-\delta)}}{64\pi m_{\chi_1}^2\left[{m_A^2}/{(4m_{\chi_1}^2)} -1 \right]^2}+ \frac{g_1^{\prime2}g_2^{\prime2}\left[\delta(2-\delta)\right]^{3/2}}{128\pi m_{\chi_1}^2\left[{m_S^2}/{(4m_{\chi_1}^2)}- 1 \right]^2}v^2 \,,
\end{align}
where there is no interference between the two mediators, and the proportionality to $\sqrt{\delta}$ indicates the available limited final state phase space. Apparently, in the case of $g'_j = g_j$ and $m_S = m_A$, the $s$-wave component mediated by $A$ dominates the DM annihilation in the non-relativistic regime.

\begin{figure}[t]
\centering
\includegraphics[width=\textwidth, trim=0 300 25 300, clip]{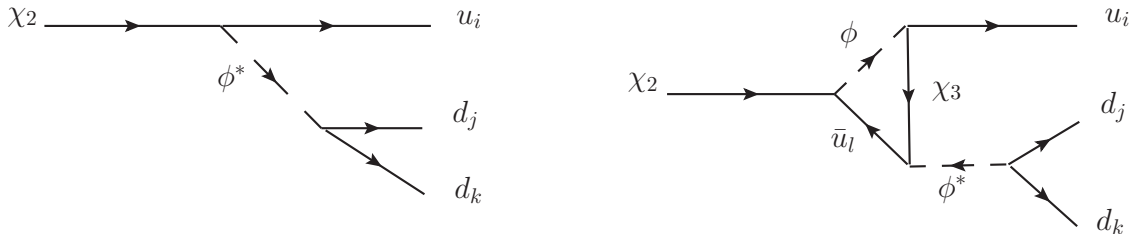}~
\caption{Tree-level and one-loop diagrams contributing to CP violation in $\chi_2$ decays.}
\label{fig:cpv}
\end{figure}
We now discuss the part that enables baryogenesis. The associated Lagrangian is given by
\begin{align}
\mathcal{L}_\text{baryog.}& = -\alpha_j\bar{u}_jP_L\chi_2\phi-\beta_j\bar{u}_jP_L\chi_3\phi-\eta_{kl}\epsilon_{kl}\phi^* \bar{d}_kP_Ld^c_l+\text{h.c.} \,, \label{eq:bg} 
\end{align}
where %
the chiral projectors 
ensure that only the right-handed quarks $u_j$ and $d_j$ of generation~$j$ participate, $\epsilon_{kl}$ is an antisymmetric symbol for flavor indices $k,l$. As previewed in general discussions, $\chi_1$ is charged under a $\mathcal{Z}_2$ symmetry. Hence, it does not couple to quarks and plays the role of the DM. $\chi_2$ is the meta-stable state whose decays trigger baryogenesis. We include a third member in the dark sector, $\chi_3$, which enables the CP asymmetry-inducing interference between tree and loop-level diagrams. A complex scalar, or ``di-quark''~$\phi$ is also introduced. This state transforms as an anti-triplet under $SU(3)_C$, has the same SM charge as the right-handed up-type quarks, and couples to the right-handed down-type quarks of the SM. All Yukawa couplings $\alpha_j,\ \beta_j,\ \eta_{jk}$ are generic, complex numbers.

Because of stringent LHC constraints on color-charged particles~$\phi$ and to allow for a tractable exposition, we shall consider the hierarchy $m_\phi \gg m_{\chi_j}$. In addition, we take $m_{\chi_2}>m_{\chi_3}$ so that contributions to the baryon asymmetry solely arises from $\chi_2$ decays as shown in Fig.~\ref{fig:cpv}. 
The analogous diagram for $\chi_3$ decay with $\chi_2$ in the loop does not contribute to an analogous CP asymmetry from $\chi_3$, since here the kinematic cut is forbidden as $m_{\chi_2} >m_{\chi_3}$.\footnote{If $m_{\chi_2} < m_{\chi_3}$, it would be $\chi_3$, instead of $\chi_2$, whose decay contributes to the CP-asymmetry by switching the subscripts $2\leftrightarrow 3$.   Alternatively, if $m_{\chi_j}\gtrsim m_\phi$, the kinematic cut can be made through $\phi$ and up-quark propagators, where both $\chi_2$ and $\chi_3$ decays would have non-vanishing $\epsilon_{CP}$, and, in principle, contribute to the baryon asymmetry. }
Finally, an ordering $|\alpha| \ll |\beta |< |\eta| $ will be assumed to ensure the branching ratio Br$_{\phi\to d_k d_l}\approx 1$, and  a sufficiently early decoupling of dark particles from the SM.   For instance and as shown below, the smallness of $|\alpha|$ together with a heavy $\phi$ guarantees that  $\chi_{1,2}$ decouple from both SM particles and $\chi_{3}$ before DM freeze-out. The hierarchy $|\alpha| \ll |\beta |$ further forces  $\chi_3$ to have a lifetime shorter than $\chi_2$, decaying into SM quarks well before BBN.     

The Lagrangian~\eqref{eq:bg} enables $\chi_2$ decay that triggers baryogenesis. With the assumed mass hierarchy $m_\phi\gg m_{\chi_j}$, we may integrate out the heavier di-quark scalar, yielding the effective operators
\begin{equation}\label{eq:decayopeff}
\mathcal{L}_{|\Delta B|=1}\supset\frac{\alpha_j\eta_{kl}}{m_{\phi}^2}\,(\bar{u}_jP_L\chi_2)\,(\bar{d}_kP_L d_l^c) + \frac{\beta_j\eta_{kl}}{m_{\phi}^2}\,(\bar{u}_jP_L\chi_3)\,(\bar{d}_kP_L d_l^c) +\text{h.c.}\ .
\end{equation}
They induce the three-body decay $\chi_{2}\rightarrow udd$ with $|\Delta B|=1$. 
The associated decay rate at $T< m_{\chi_2}$ is given by
\begin{align}\label{eq:chi2dec}
  \Gamma_{\chi_2,\text{dec}}\approx\frac{3|\alpha_j|^2|\eta_{kl}|^2 m_{\chi_2}^5}{1024\pi^3m_\phi^4}\, ,
\end{align}
where the factor of three in Eq.~\eqref{eq:chi2dec} accounts for color multiplicity. The CP-asymmetry produced in decays arises through the interference between tree-level and 1-loop diagrams shown in Fig.~\ref{fig:cpv}. The asymmetry as a function of $m_{\chi_2}$, $m_{\phi}$, and Yukawa couplings is given by~\cite{Cui:2013bta}
\begin{equation}\label{eq:cpv}
\epsilon_{CP}=\text{Im}[\alpha_j\alpha_m\beta_j^*\beta_m^*]\frac{ m_{\chi_2}^2 }{20\pi|\alpha_j|^2 m_\phi^2 }\,,
\end{equation} 
in the process of $\chi_2 \to u_j d_k d_l$ (and $\bar u_j \bar d_k \bar d_l$) with an intermediate up-type quark $u_m$.
Since we are interested in a broad range of $m_{\chi_1}$ spanning over $\mathcal{O}(10-10^4)~\text{GeV}$, all generations may contribute when $m_{\chi_2}\gtrsim m_t$. Nevertheless, the constraints from (di-)nucleon decay and neutron-antineutron oscillation on the effective operators in Eq.~\eqref{eq:decayopeff} are very strong if $(u_j d_k d_l)$ that couples to $\chi_3$ contains only light quarks, such as $(uds)$; see e.g.,~\cite{Aitken:2017wie}. Thus, here we focus on the combination of $(cds)$, or $(udb)$, where sizable coupling-combination $\beta \eta$ is allowed, and $\chi_{2,3}$ decays into them are kinematically accessible. 
Finally, we note that the model is exempt from constraints from the null measurements of the neutron electric dipole moment (EDM). The reason is that the interference diagrams in Fig.~\ref{fig:cpv} leading to CP violation only involve right-handed quarks~\cite{He:1996hb, Yamanaka:2017mef}.

 \begin{figure}[tb]
\centering
~\includegraphics[width=0.5\textwidth, trim=0 0 0 0, clip]{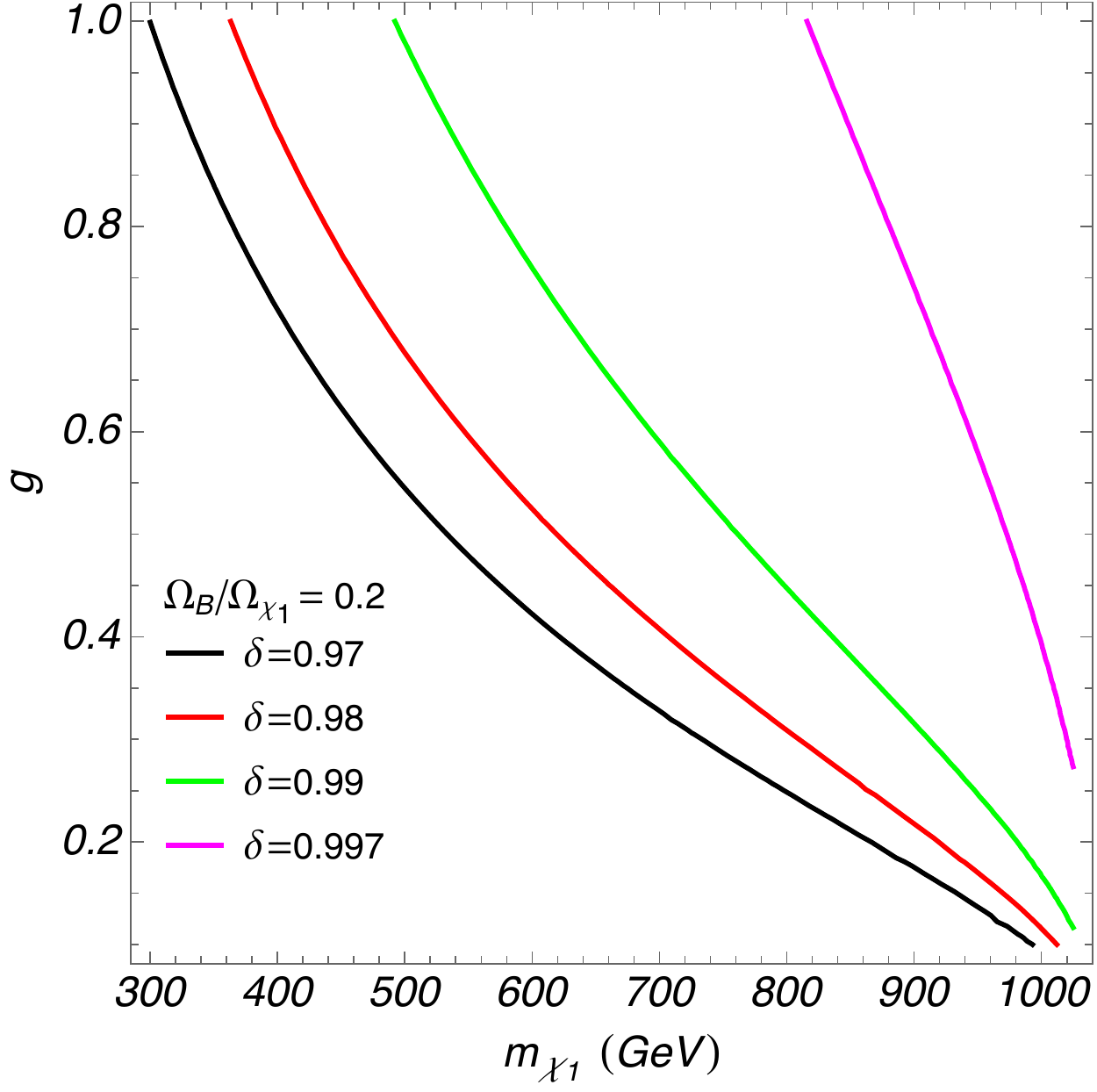}~
~\includegraphics[width=0.5\textwidth, trim=0 0 0 0, clip]{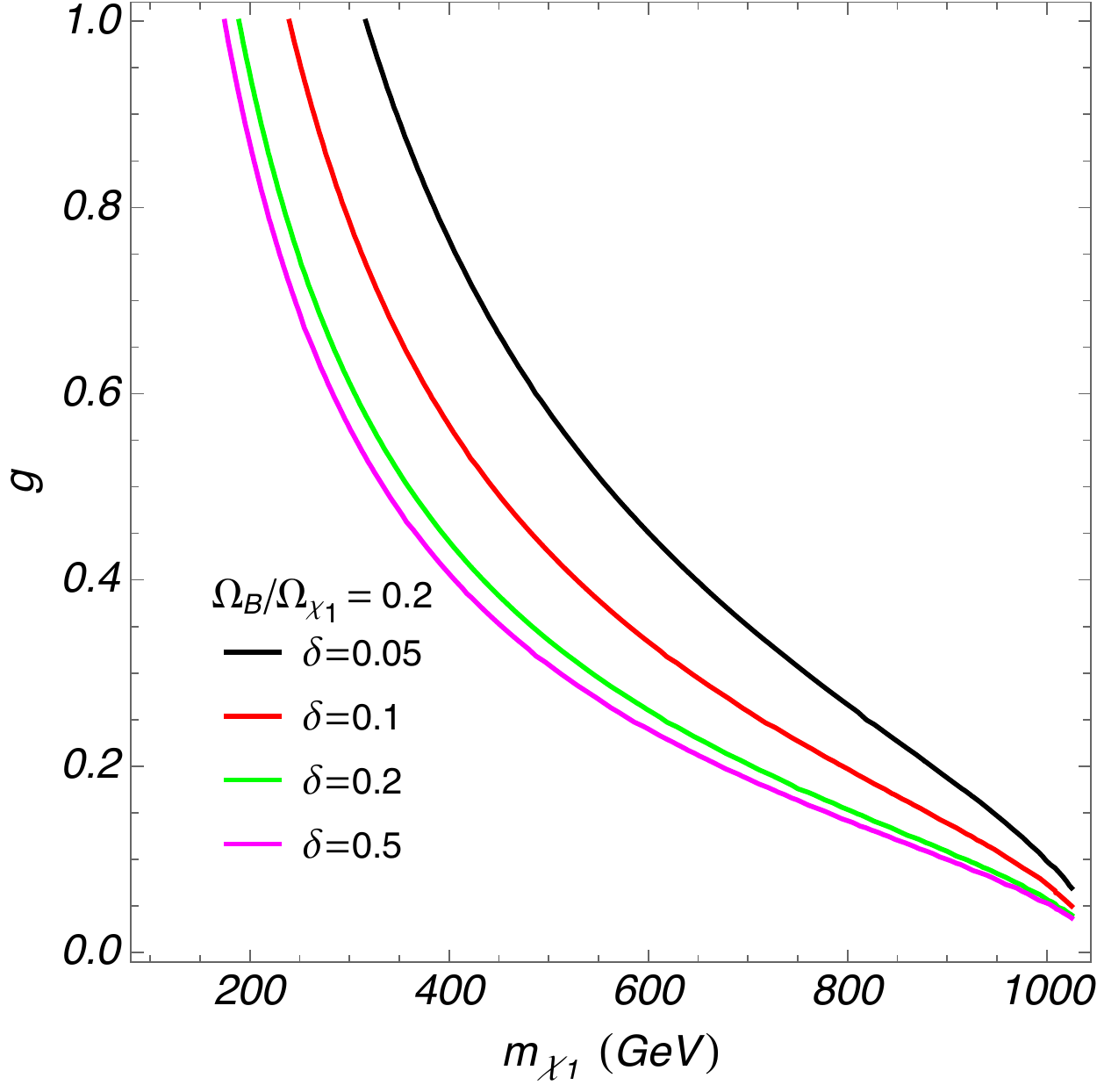}
\caption{Contours of constant $\Omega_B/\Omega_{\chi_1}= 0.2$ for varying annihilation coupling $g$ and DM mass $m_{\chi_1}$ and four values of the mass splitting $\delta$ in the hierarchical (left) and the nearly degenerate (right) scenarios where annihilation is dominantly $s$-wave with $g\equiv \sqrt{g_1 g_2}$ and/or $m_A\lesssim m_S$. The other model parameters are pseudoscalar mass $m_A=2.1~\text{TeV}$, $\phi$ mass $m_{\phi}=10~\text{TeV}$, initial temperature ratio $\xi_i=0.1$, and $\chi_3-\text{quark}$ Yukawa coupling $\beta=1$.} 
\label{fig:Brat}
\end{figure}

Having established all the ingredients and hierarchies of a UV complete model, we now turn to the identification of prospective parameter ranges. 
As discussed above, DM decoupling with $\xi_i \le 1$ prefers an earlier freeze-out in comparison with the standard case. Under the assumption that $\chi_1$ freezes out for $T\gtrsim  m_{\chi_2}/10$, the requirement that $\chi_2$ decays after $\chi_1$ freeze-out but before BBN yields the preferred range of allowed decay couplings as
\begin{equation}\label{eq:yukconstraint}
 10^{-9} \left({1\, \text{TeV}\over m_{\chi_2}} \right)^{5/2} \left({ m_\phi  \over 10\, \text{TeV} } \right)^2 \le  |\alpha | |\eta| \le 10^{-4} \left({1\, \text{TeV}\over m_{\chi_2}} \right)^{3/2} \left({ m_\phi  \over 10\, \text{TeV} } \right)^{2} \,.
\end{equation}
 Note that the last inequality is also approximately the condition that $\chi_{2}$ decouples from the SM bath during  $\chi_1$ freeze-out: if the bound were violated, the process $\chi_2 \leftrightarrow u_j d_k d_l$ becomes efficient and dark and SM sectors thermalize. As a result, for $ m_\phi = 10$\,TeV, the hierarchy $|\alpha| \ll |\eta|$ introduced for baryogenesis leads to $|\alpha| \ll 10^{-2}$ for $m_{\chi_2} = 1\,$TeV, and $|\alpha| \ll 0.14$ for $m_{\chi_2} =30\,$GeV.  Similarly, requiring $\chi_3$ to decay fast, say at $T \sim m_{\chi_3} $, suggests $|\beta||\eta| \gtrsim 0.1 (m_\phi/10\,\text{TeV})^4 (10\,\text{GeV}/m_{\chi_3} )^3$, which can be easily satisfied in the parameter region of our interest here. So the existence of $\chi_3$ neither leads to additional early matter domination, nor affects the standard BBN processes.

\begin{figure}[tb]
\centering
~\includegraphics[width=0.5\textwidth, trim=0 0 0 0, clip]{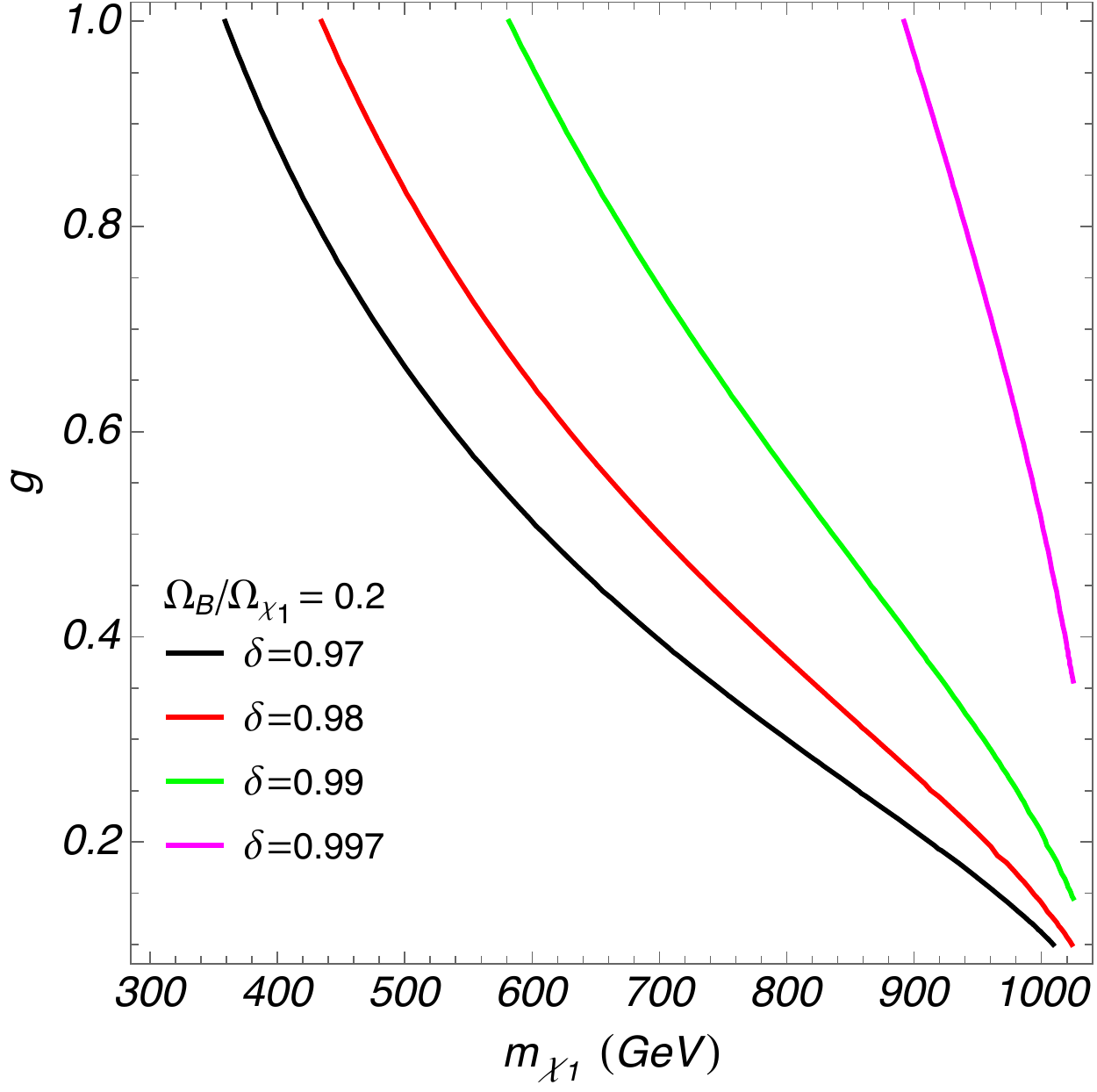}~
~\includegraphics[width=0.5\textwidth, trim=0 0 0 0, clip]{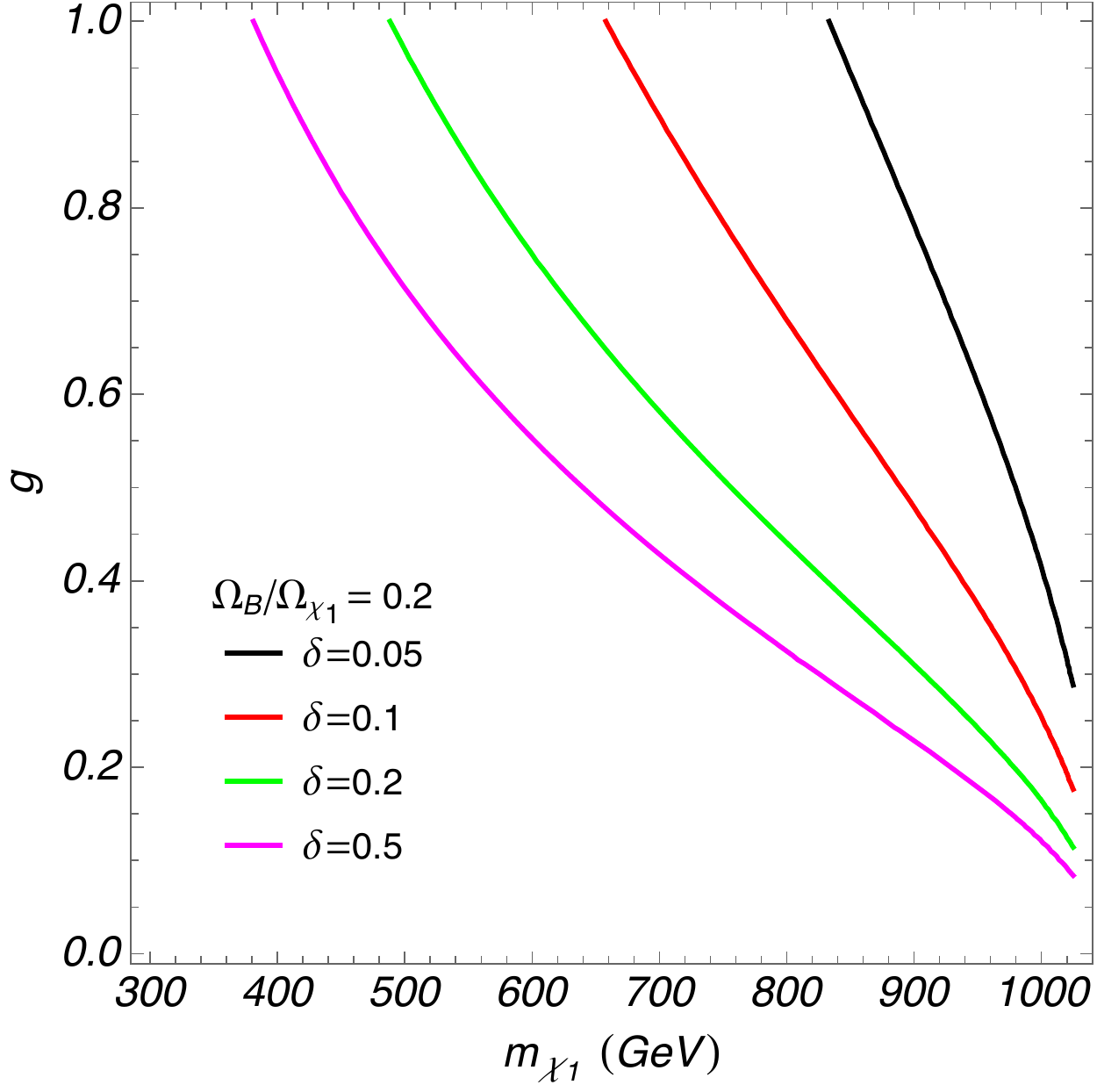}

\caption{Contours of constant $\Omega_B/\Omega_{\chi_1}\approx 0.2$ for varying annihilation coupling $g$ and DM mass $m_{\chi_1}$ and four values of the mass splitting $\delta$ in the hierarchical (left) and the nearly degenerate (right) scenarios where annihilation is dominantly $p$-wave in the limit of $m_{A}\gg m_S$. In analogy to the choice made in Fig.~\ref{fig:Brat}, we choose $g^{'}\equiv \sqrt{g^{'}_1 g^{'}_2}$. The other model parameters are  $m_S=2.1~\text{TeV}$,  $m_{\phi}=10~\text{TeV}$, initial temperature ratio $\xi_i=0.1$, and $\chi_3$-$\text{quark}$ Yukawa coupling $\beta=1$.} 
\label{fig:Bratpwave}
\end{figure}

Putting all together and plugging the specific results into Eq.~\eqref{eq:coincidence}, we obtain the relic abundance of DM and the baryon asymmetry. Taking Eq.~\eqref{eq:swave}, their ratio in the exemplary UV model for $s$-wave dominated freeze-out becomes 
\begin{equation}
\frac{\Omega_{B}}{\Omega_{\chi_1}} \sim 0.2 \left({|\beta|^2 \sin2\theta_\beta \over x_{\rm f.o.}}\right) \left({\xi_i \over 0.1}\right)^3 \left({\sqrt{g_1 g_2} \over 0.2}\right)^4 \left({30\,{\rm TeV} \over m_\phi  (  m_A^2 / 4m_{\chi_1}^2 -1) }\right)^2 \left({(1-\delta)^2 \sqrt{\delta} \over 0.1}\right) \,, 
\end{equation}
for which, and for numerical results below, we have assumed the Yukawa couplings between $\chi_2$ and the up-type quarks to be real such that the CP-asymmetry in Eq.~\eqref{eq:cpv} can be written in terms of a complex phase $\theta_\beta$ of $\beta:~\text{Im}[(\alpha\alpha\beta^*\beta^*)^2]\rightarrow|\alpha|^2|\beta|^2 \sin(2\theta_\beta)$. 
We observe, that the final ratio of $\Omega_{B}$ to $\Omega_{\chi_1}$ is sensitive to the mass splitting $\delta$ and $m_A /m_{\chi_1}$, instead of $m_{\chi_1}$ alone. 
In Fig.~\ref{fig:Brat} we show contours of $\Omega_B /\Omega_{\chi_1}= 0.2$ as a function of the $\chi_1$ mass and DM annihilation coupling $g \equiv \sqrt{g_1 g_2}$ %
in the hierarchical (left) and the nearly degenerate (right) cases of our exemplary model. Note that the contributions from the scalar $S$ is subleading and is thus not taken into account. To obtain the numerical results, we plug Eqs.~\eqref{eq:sannihilation} and \eqref{eq:cpv} into the solution for the $\chi_1$ freeze-out abundance given by Eqs.~\eqref{eq:chi1ab} and \eqref{eq:chiabund}, which is then used together with $Y_{\chi_2, f}\approx\left(0.42\xi_i^3 {g_{\chi}}/{g_{*}}-Y_{\chi_1,f}\right)$ to find the abundance of $\chi_2$ just after DM freeze-out. This, in turn, gives rise to the observed baryon asymmetry, $Y_{B} \propto \epsilon_{CP}Y_{\chi_2,f}$. The contours have fixed mass splitting in the range of $0.97<\delta<0.997$ (or equivalently, $0.003 m_{\chi_1}< m_{\chi_2} < 0.03 m_{\chi_1}$) in the hierarchical case and $0.05<\delta<0.5$ in the degenerate case. The pseudoscalar mass is taken to be $m_A=2.1~\text{TeV}$ in Fig. \ref{fig:Brat}. In the limit of $m_A \gg m_S $ and/or $g^{'}\gg g$, the annihilation channel via the scalar $S$ dominates the DM freeze-out, which case is shown in Fig. \ref{fig:Bratpwave}. In our analysis, we fix $\theta_\beta= \pi/4$ to obtain the maximal CP-asymmetry.

\section{Experimental Signatures}\label{sec:signatures}

The concrete model outlined in the previous section lends itself to 
observational signatures and experimental tests, in particular in relation to the stable DM candidate $\chi_1$ and its unstable dark sector partners $\chi_{2,3}$. These include conventional and novel direct and indirect detection signals, as well as modifications to primordial density fluctuations. We will sketch the prospects of these searches below, leaving a more-in-detail analysis for future work. 

\subsection{New particle searches at colliders}
\begin{figure}[h!]
\centering
\includegraphics[width=0.4\textwidth, trim=10 140 0 140, clip]{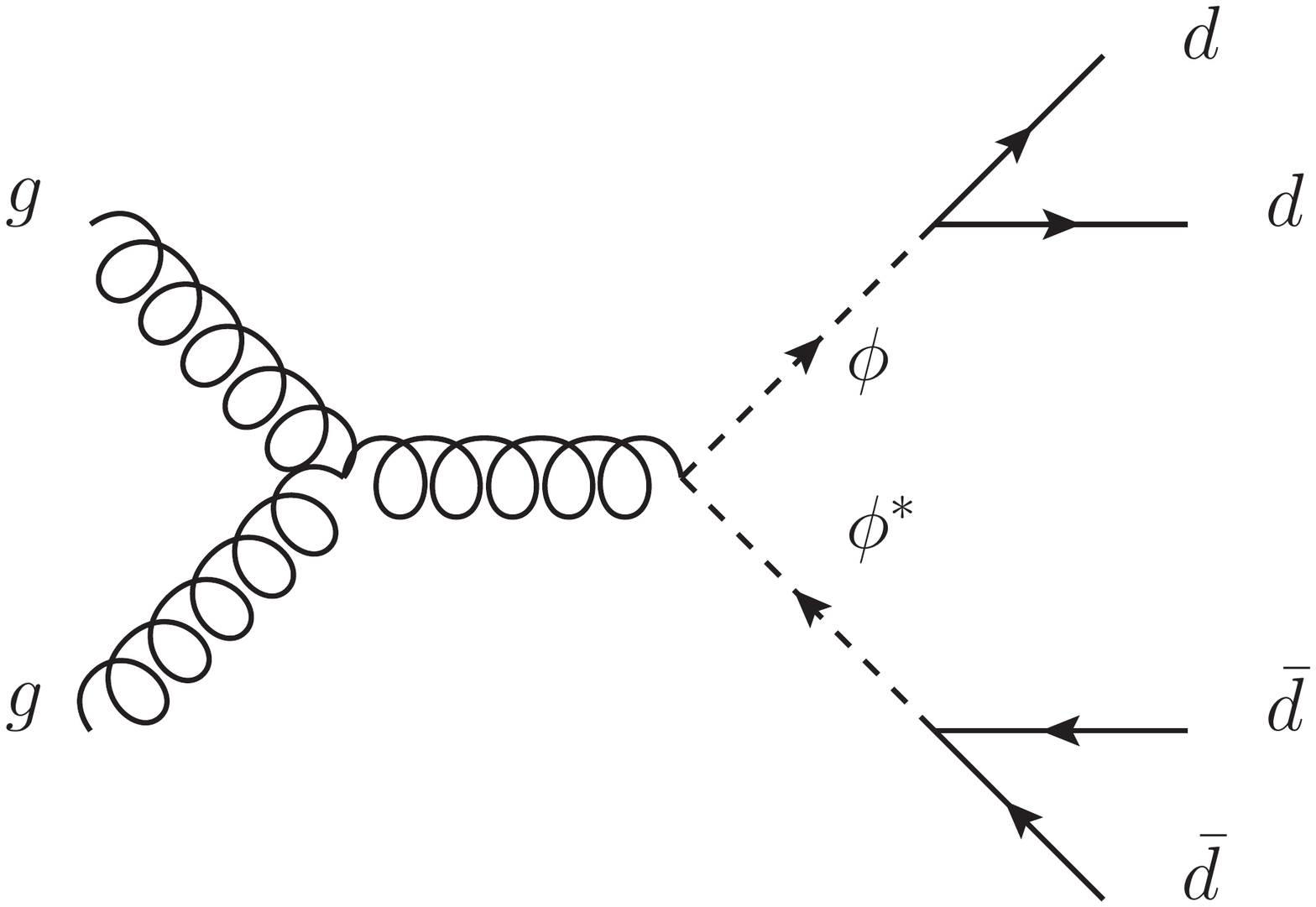}~
\includegraphics[width=0.4\textwidth, trim=10 140 0 140, clip]{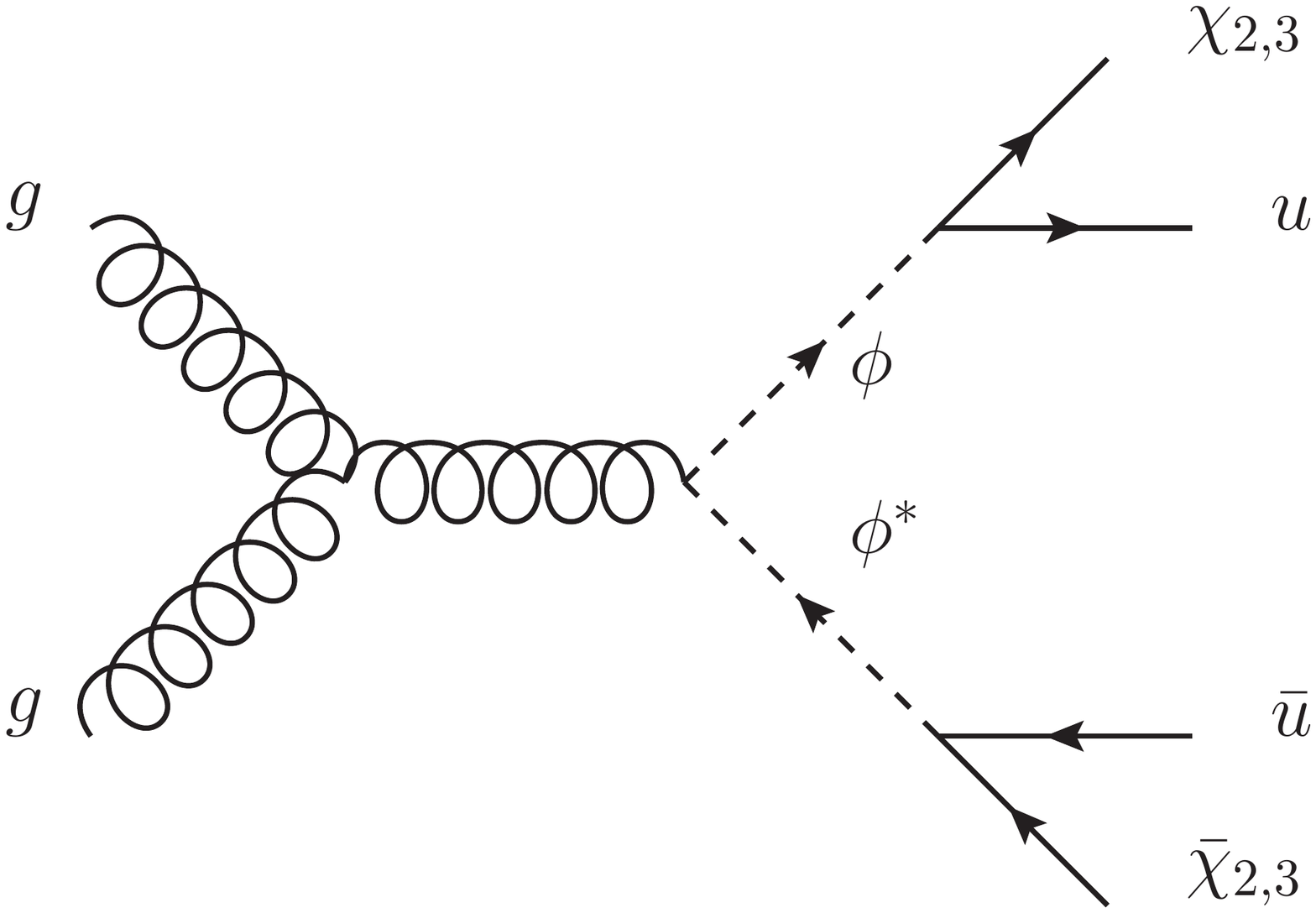}
\caption{Diagrams relevant for $\phi$ (left) searches and $\chi_{2,3}$ production at the LHC. }
\label{fig:colliderphi}
\end{figure}
To produce the baryon asymmetry with renormalizable operators, the color anti-triplet, di-quark scalar $\phi$ has been introduced. As a result of the color charge it carries, LHC bounds on its mass are quite strong. As outlined in Sec.~\ref{sec:model}, the di-quark scalar $\phi$ decays to two down-type quarks. At the LHC, $\phi\phi^*$ are pair-produced dominantly via gluon fusion, and subsequently decay to 4 jets: $pp\rightarrow\phi\phi^*\rightarrow4j$, as shown in Fig.~\ref{fig:colliderphi}. LHC searches for di-quark scalars constrain the mass of $\phi$. For instance, the recent CMS search~\cite{CMS:2018mgb} constrains the production cross section of di-quark scalar resonances to be below fb at $\sqrt{s} =13$\,TeV, excluding $m_{\phi}\lesssim 7~\text{TeV}$. %

\begin{figure}
\centering
\includegraphics[width=0.4\textwidth, trim=10 140 0 140, clip]{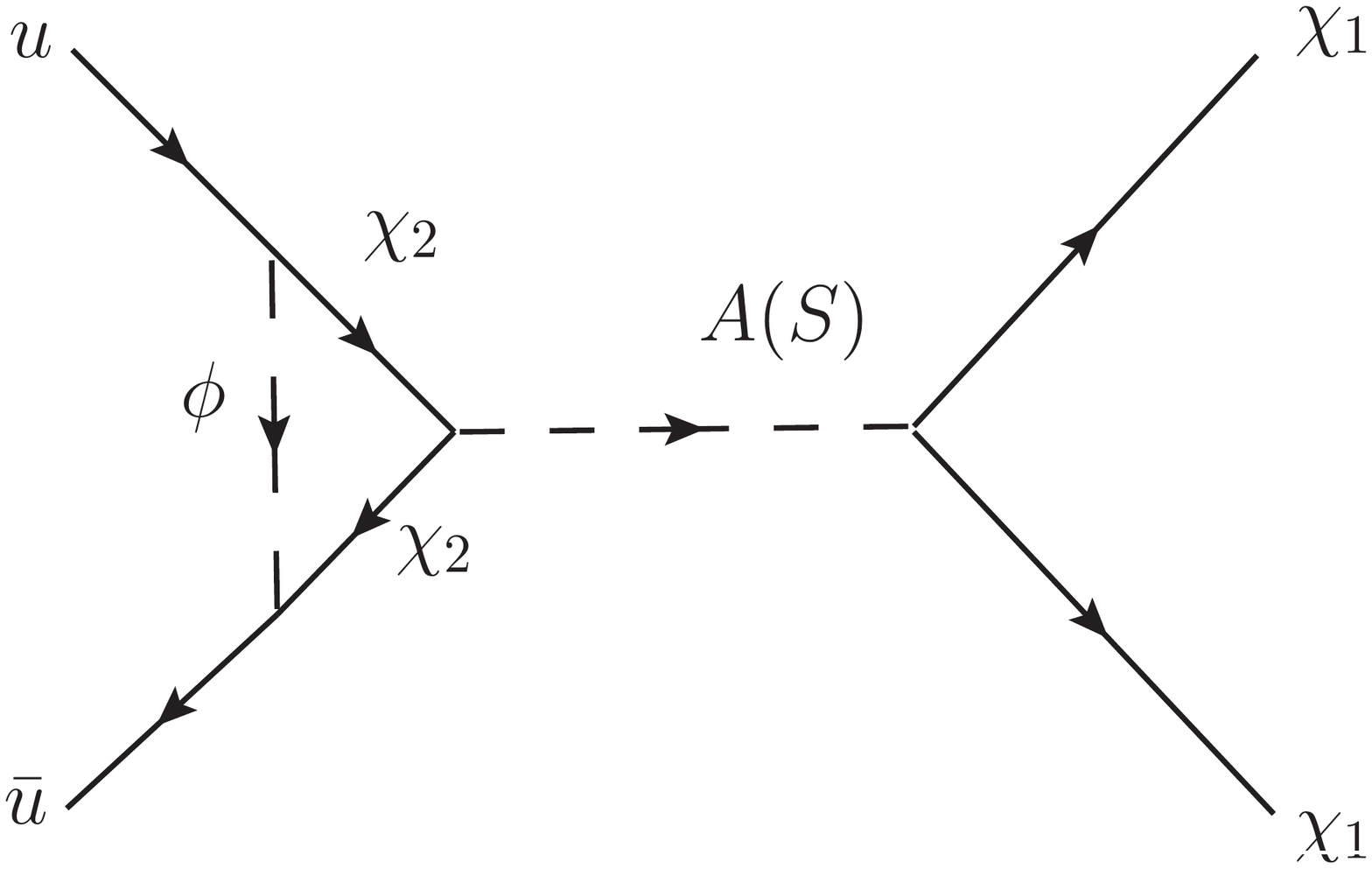}~
\includegraphics[width=0.35\textwidth, trim=0 100 10 100, clip]{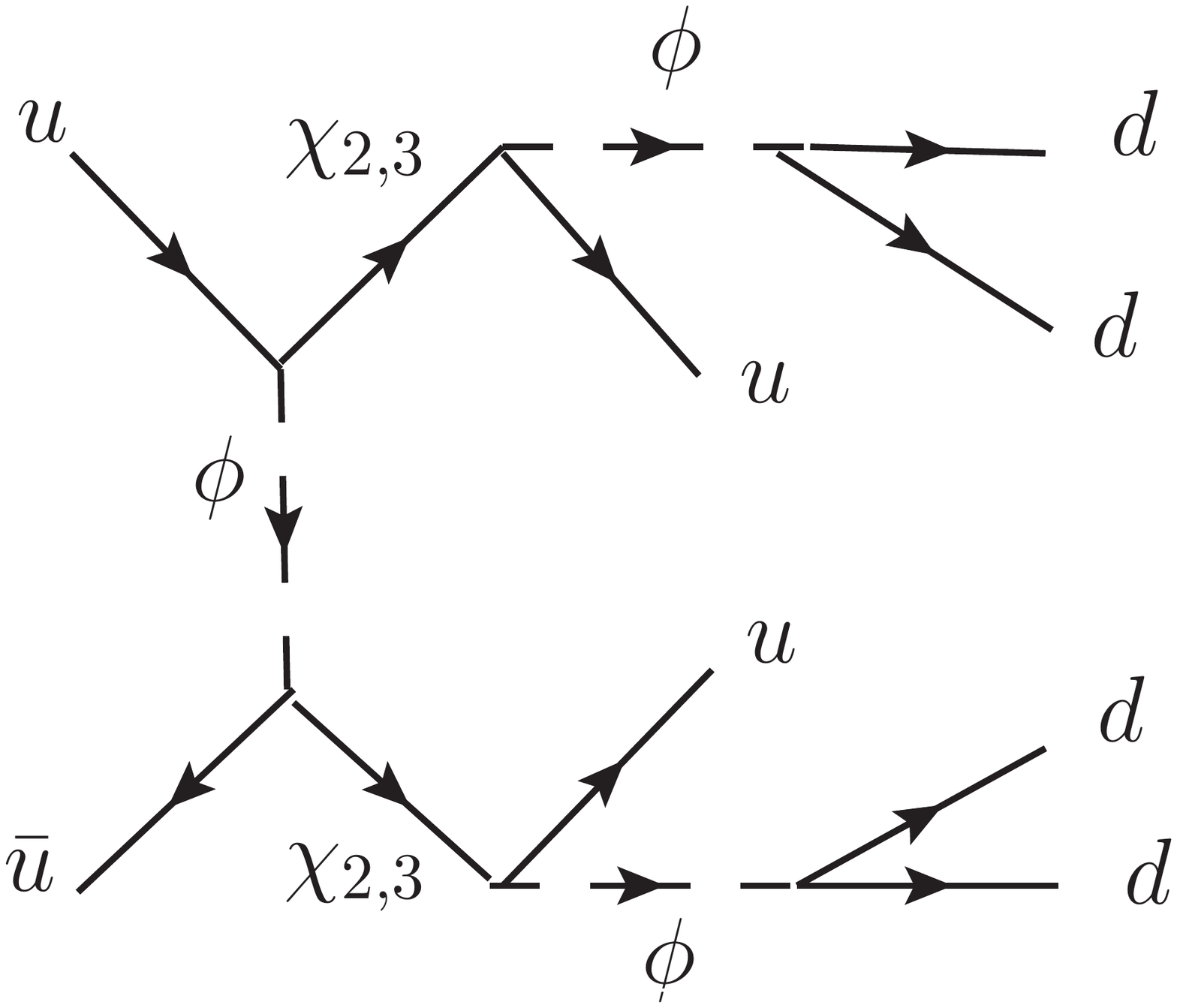}
\caption{Diagrams showing dominant contributions to $\chi_1$, $\chi_2$ and $\chi_3$ production at the LHC. %
}
\label{fig:colliderchi}
\end{figure}

The $\chi_j$ ($j=1,2,3$) particles can also potentially be produced at colliders.
The production of $\chi_1$ is shown on the left panel of Fig.~\ref{fig:colliderchi}, and would lead to a missing transverse energy (MET) signal. However, due to the combination of loop factor, the required large mass of the di-quark scalar $\phi$ in the loop diagram, and the additional initial state radiation (ISR) photon or jet required for tagging the event, the $\chi_1$ production rate is found to be too small to be detectable. The production of $\chi_2$ and $\chi_3$ can proceed via t-channel $\phi$-exchange as shown in the right panel of Fig. \ref{fig:colliderchi}. The subsequent $\chi_{2,3}$ decays can lead to prompt jets, displaced vertices or MET, depending on their decay length. Again, $m_\phi\gtrsim 7$~TeV suppresses the production rates of $\chi_2$ and $\chi_3$. Since successful co-genesis requires $|\alpha|\ll 10^{-2}$ for $m_{\chi_2}=1$~TeV, the production cross section of $\chi_2$  falls significantly below fb, well beyond the LHC reach. The prospect for $\chi_3$ production is better, as the coupling $|\beta|$ can be $O(1)$, leading to fb-sized LHC production cross sections for $\sqrt{s}= 13\,$TeV and $m_{\phi} = 10\,$TeV~\cite{Kilic:2015vka}. The prospect of detecting $\chi_3$ depends on its decay length following production. It may vary over a wide range,
\begin{equation}
  c\gamma \tau_{\chi_3} = 4 \text{\,cm} \left({\gamma \over  10 } \right) \left({10^{-4} \over \beta \eta } \right)^{2} \left({1\, \text{TeV}\over m_{\chi_3}} \right)^{5} \left({ m_\phi  \over 10\, \text{TeV} } \right)^4 \,,
\end{equation}
where $\gamma = E_{\chi_3}/m_{\chi_3}$ is the Lorentz factor; the same expression also applies to $\chi_2$ after replacing $m_{\chi_3}$ and $\beta$ with $m_{\chi_2}$ and $\alpha$. $\chi_3$ decay can hence lead to prompt multijet events, MET, or displaced jet signals. For the size of the production cross section, the displaced vertex channel is the most promising one, with the potential of detection at the High-Luminosity LHC \cite{Cui:2014twa}, due to the generally low background \cite{CMS:2019zmd,CMS:2018tuo, CidVidal:2018eel, ATLAS:2018yii}.

We would like to note that although with the LHC it is generally challenging to detect $\chi_j$ particles in the model as outlined above, the detection prospects are expected to improve notably with the proposed future higher energy colliders \cite{Mangano:2017tke, ILC:2019gyn, CEPCStudyGroup:2018ghi, Roloff:2018dqu, FCC:2018byv}. Furthermore, model-variations can increase the production rate of $\chi_j$. 
For instance, interaction terms, such as a trilinear $A$-$A$-$H$, may open up new channels at higher energy colliders for both $\chi_1$ and $\chi_2$ production with potentially appreciable rate, independently of $m_\phi$.
In addition, while we consider $\phi$ dominantly decays into a pair of down-type quarks, a sizable branching ratio of $\phi \to \chi_3 u $ is possible, depending on the ratio $|\beta/\eta|^2$, which provides a new, potentially efficient channel for $\chi_3$ production as shown in the right panel of Fig.~\ref{fig:colliderphi}; for $\chi_2$ the corresponding ratio is much smaller and less prospective. 

In summary, the di-quark and $\chi_{j\,}$'s in the co-genesis framework that utilizes quark-couplings in its UV-complete representation could leave observable signals in various search channels at high luminosity run of the LHC, yet are potentially challenging. On the other hand, the proposed future colliders at the high energy frontier have more promising capacity to reveal detectable signatures from this framework.

\subsection[Direct detection of $\chi_1$]{Direct detection of \boldmath$\chi_1$}
\label{sec:DD}

\begin{figure}[h!]
\centering
~\includegraphics[width=0.4\textwidth, trim=0 150 50 150, clip]{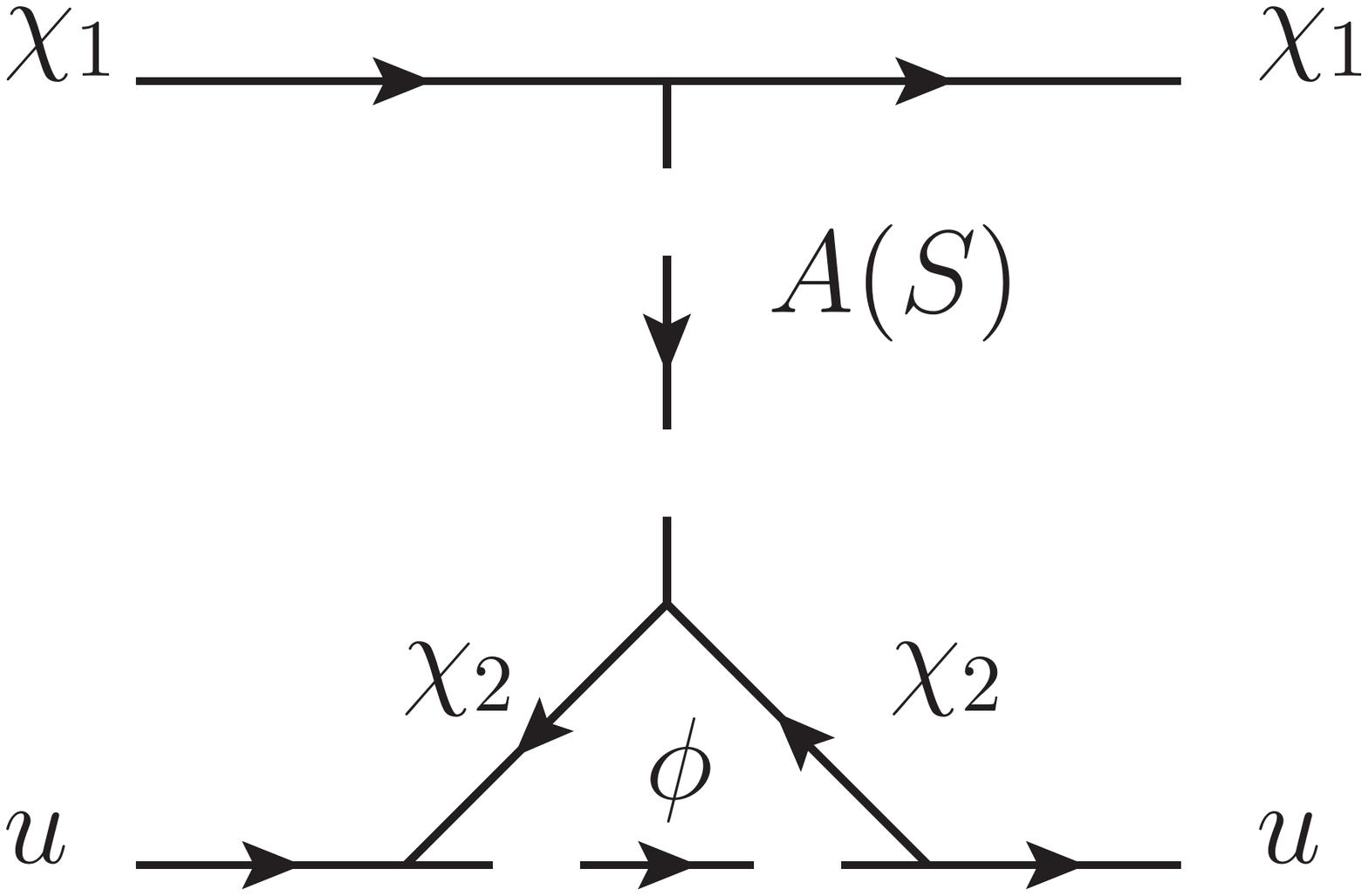}~
~\includegraphics[width=0.4\textwidth,trim=0 150 50 150 clip]{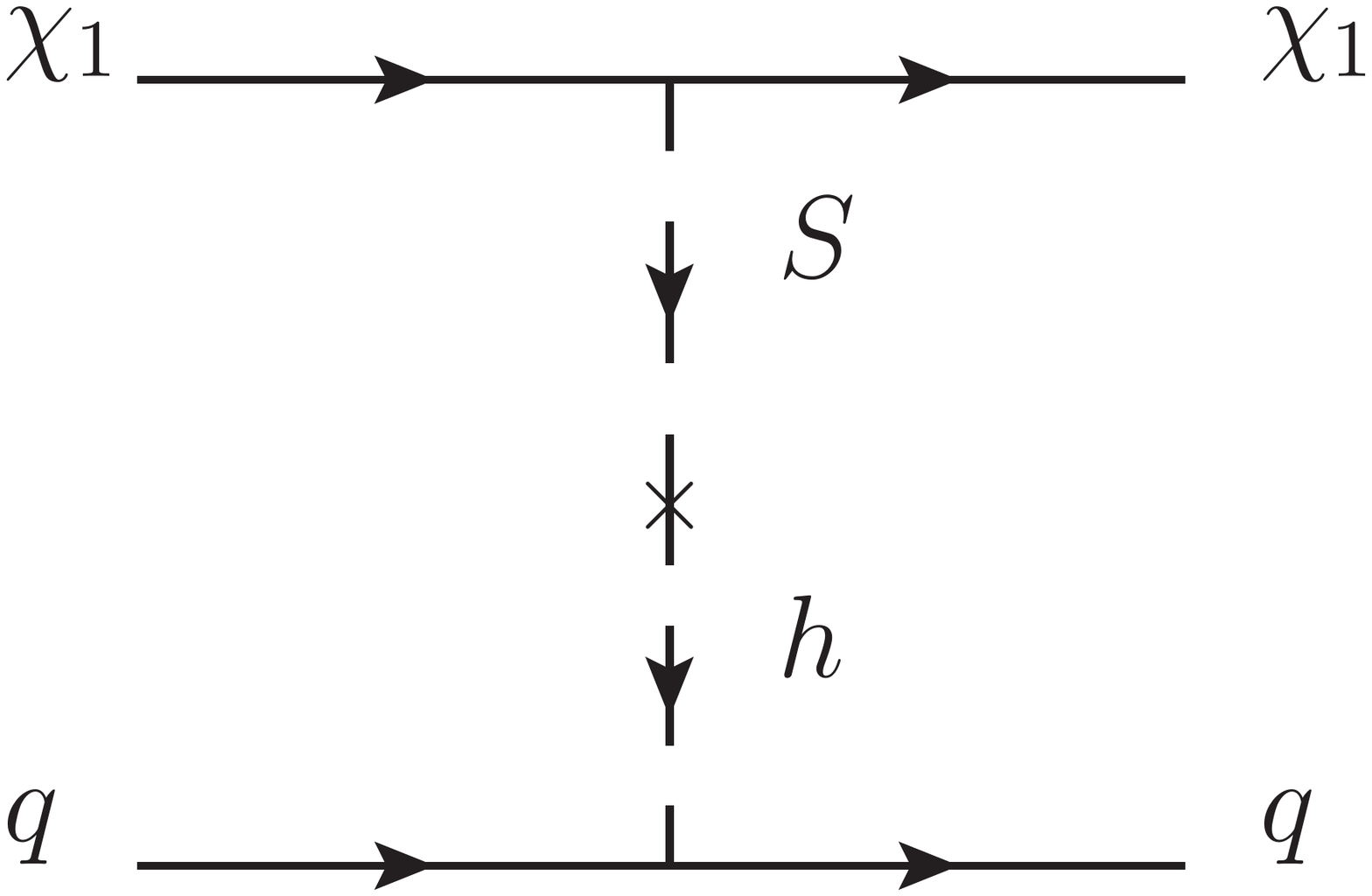}
\caption{Diagrams contributing to DM-nucleus scattering $\chi_1N\rightarrow\chi_1N$. The diagram on the left is the dominant contribution in the minimal model while the diagram on the right may dominate if the $S$-Higgs mixing is present. %
}
\label{fig:dd}
\end{figure}

In the example model detailed in Sec.~\ref{sec:model}, there are no interactions between the stable DM candidate $\chi_1$ and quarks at tree-level. However, DM-nucleus interactions can be induced at 1-loop as shown in the left panel of Fig.~\ref{fig:dd}. Integrating out the heavier mediating states generates low-energy effective operators which induce interactions between $\chi_1$ and nucleons $N=n,\ p$. They are either spin-independent, $\mathcal{L}\propto\bar{N}N S$, or spin-dependent, $\mathcal{L}\propto\bar{N}(i\gamma^5)N A $. 
The effective interactions lead to  DM-nucleon elastic scattering cross sections which can be estimated using dimensional analysis: 

\begin{equation}\label{eq:dd}
\sigma^{\rm SI}_{\chi_1N}\sim {f_N^2\over 256\pi^5}\left[\frac{\alpha^2g'_1g'_2m_{\chi_2} m_{\chi_1}m_N^2}{m_{\phi}^2m_S^2(m_{\chi_1}+m_N)}\right]^2 \text{~and~}\, \sigma^{\rm SD}_{\chi_1N}\sim \frac{v^2 f_N^2}{512\pi^5}\left[\frac{\alpha^2g_1g_2m_{\chi_2}^2m_{\chi_1}m_N}{m_{\phi}^2m_A^2(m_{\chi_1}+m_N)}\right]^2 ,
\end{equation}
which applies in the limit $m_{\phi, S}\gg m_{\chi_{2} } \gg m_N$~\cite{Kumar:2013hfa}. Here the first component, mediated by $S$, is a spin-independent cross section, while the latter, mediated by $A$, is both velocity suppressed and spin-dependent; the coefficient $f_N \approx 0.3$, is counting for the valence quark content of the nucleon~\cite{COMPASS:2007esq}. 
For instance, under the assumption that $g_1 g_2/m_A^2 = g'_1 g'_2/m_S^2 =1/(4\,\text{TeV}^2)$, $m_{\chi_1} =3\, $TeV, and $m_{\chi_2} =0.1\, $TeV, which gives the observed DM abundance with $\xi\sim 1$, setting $m_{\phi} = 10\,$TeV together with Eq.~\eqref{eq:yukconstraint} require $|\alpha| \ll 1$, resulting in the spin-independent part
$
\sigma^{\rm SI}_{\chi_1N} \ll  10^{-56}\,\text{cm}^2\,
$;
the spin-dependent component is even smaller due to the velocity suppression. We conclude that in the current setup, there are no direct detection prospects. 
Finally, we point out that there can be contributions to direct detection at tree-level when the scalar $S$ mixes with the SM Higgs. In order for the dark sector and SM to remain decoupled as per our working hypotheses, the $S$-Higgs mixing angle must be small, $\theta_S \lesssim10^{-7}$. In the case of $m_S \ge m_h$, the scattering cross section between $\chi_1$ and proton through mixing can be estimated as $ 10^{-42} ( {g'_1 \theta_S } )^2 \,\text{cm}^2$~\cite{Lopez-Honorez:2012tov}, and thus will not be detectable in the foreseeable future.

\begin{figure}[h!]
\centering
~\includegraphics[width=0.5\textwidth, trim=0 140 10 140, clip]{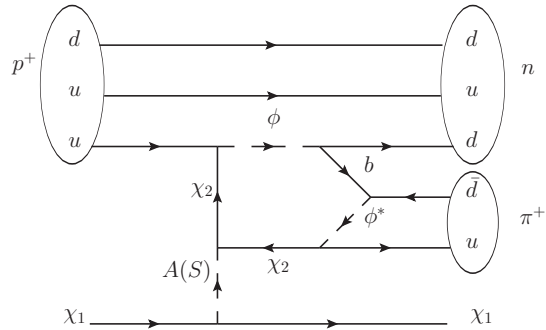}~
\caption{Example of exotic nucleon conversion arising in dark sector baryogenesis. %
}
\label{fig:indfig}
\end{figure}

An interesting possibility arises through $t$-channel scattering of protons (or neutrons) with $\chi_j$ at one loop via off-shell $A\,(S),~\chi_2,~\text{and}~\phi$. An example of this process with a proton initial state is shown in Fig.~\ref{fig:indfig}. 
Therefore, it is important to note that the relative kinetic energy of the galactic $\chi_1$ DM flux at earth interacting with nuclei falls short to kinematically access this channel, but rather requires a ``boosted'' $\chi_1$ component. 
A particularly interesting environment may involve neutron stars in regions of high $\chi_1$ density accelerating $\chi_1$ to sufficiently large energy~\cite{Guver:2012ba}, such that this nucleon conversion process occurs and leads to observable effects such as anomalous heating. Note that the energy spectrum of the incoming and outgoing states can differ significantly from the elastic scattering usually considered \cite{Acevedo:2019agu,Bell:2018pkk}. We leave a detailed investigation of nucleon conversion via capture in neutron stars, and other stellar objects, to future work.

\subsection[(Exotic) indirect detection through $\chi_2$]{(Exotic) indirect detection through \boldmath$\chi_2$}

The low-redshift $\chi_1$ annihilation into possibly energetic $\chi_2$ particles shares some features with the boosted dark matter scenario~\cite{Agashe:2014yua, Berger:2014sqa}. However, the $\chi_2$ final states decay electromagnetically within a second, leading to box-shaped energy spectra of final SM quarks~\cite{Ibarra:2012dw}, which in turn can be experimentally constrained. Approximately, the annihilation cross section is given by $\langle \sigma_{\chi_1} v\rangle \approx (\xi_\text{f.o.}/\Delta)$\,pb, in order to yield the observed DM abundance. This leads to lower bounds on DM mass around $m_{\chi_1} \ge 50\, (\xi_\text{f.o.}/\Delta)$\,GeV from Planck and indirect search experiments (see e.g.,~\cite{Elor:2015bho, Slatyer:2015jla, Fermi-LAT:2019lyf, Kahlhoefer:2021sha})\footnote{Extremely small mass splittings, below the typical kinetic energy of galactic DM, can lead to a velocity-dependent cross section. This is a fine-tuned situation that does not happen in the generic parameter space of interest here. }. 
 
In addition, DM can be captured by celestial objects, and consequently annihilate inside. Given the short lifetime of $\chi_2$, one particularly interesting signature is that the boosted $\chi_2$ produced from DM annihilation in the centre of Earth decays inside a large-volume terrestrial detector, for which we take IceCube~\cite{IceCube:2016aga} as an example. The signal requires the in-flight decay length of $\chi_2$, $l_{\chi_2}\equiv \tau \,c \sqrt{m_{\chi_1}^2/m_{\chi_2}^2-1}$, to be on the order of or longer %
than the Earth's radius, $r_{\rm Earth}$. If satisfied, the probability for $\chi_2$ to decay inside its detector of fiducial volume $1\,$km$^3$ is given by 
\begin{equation}\label{eq:iceCubeP}
  \left. P_{\chi_2 \to X}\right|_{\text{detector}} ={V_\text{detector} \over l_{\chi_2} \cdot 4\pi r_{\rm Earth}^2 } \approx 0.8\times 10^{-13}  \left({\text{0.02\,sec} \over \tau \sqrt{m_{\chi_1}^2/m_{\chi_2}^2-1}}\right)\,.  
\end{equation}
Assuming a sensitivity threshold of several events per year in the detector same as \cite{Feng:2015hja} and $l_{\chi_2} /c=\text{0.02\,sec}$, the probability given by Eq.~\eqref{eq:iceCubeP} implies that the DM annihilation rate, $\Gamma_{A}$, in the Earth's centre should reach $10^6$ particles per second, in order to be detectable in IceCube, which is in agreement with \cite{Schuster:2009au}. 

We may consider the prospects of such signature in the concrete UV model. As discussed in Sec.~\ref{sec:DD}, the non-relativistic scattering between $\chi_1$ and nucleons is dominated by the $S$-mediated interaction, leading to a spin-independent cross section~$\sigma^{\rm SI}_{\chi_1N}$. In the parameter region of interest, where the annihilation cross section is $\langle \sigma_{\chi_1} v\rangle \le 1\,$pb, equilibrium between capture and annihilation in Earth is not reached~\cite{Lundberg:2004dn, IceCube:2016aga}. That is, the actual DM annihilation rate in Earth scales as $(\sigma^{\rm SI}_{\chi_1N})^2\, \langle \sigma_{\chi_1} v\rangle$, up to subleading effects~\cite{IceCube:2016aga}. Moreover, a comparison of Figs.~6 and~8 in \cite{IceCube:2016aga} allows us to estimate the coefficient of this scaling for $m_{\chi_1} \ge 100\,$GeV, leading to 
\begin{equation}
  \Gamma_{A} \sim  10^{9} \, \text{s}^{-1} \left({10 \,\text{TeV} \over m_{\chi_1} }\right)^{2} \left({\sigma^{\rm SI}_{\chi_1N}\over 10^{-41}\text{cm}^2}\right)^2 \left({\langle \sigma_{\chi_1} v\rangle \over 1\,\text{pb}}\right)\,.
\end{equation}
As an example, for 100\,GeV DM with $\langle \sigma_{\chi_1} v\rangle = 0. 1\,$pb, detecting $O(1)$ events per year requires $\Gamma_{A} \sim 10^6$\,s$^{-1}$, implying a sensitivity of $\sigma^{\rm SI}_{\chi_1N}\sim 10^{-45}-10^{-46}\rm cm^2$. This is comparable to current direct detection constraints, but still needs to be improved by several orders of magnitude to eventually probe our model. Such improvement may be achieved if the DM annihilation is significantly enhanced at low velocities (e.g.,~through Sommerfeld enhancement with a light mediator), bringing the Earth capture and annihilation into equilibrium. 
Note that this simple re-scaling does not apply for $m_{\chi_1} \le 100\,$GeV, where the bound gradually gets weaker, except for DM masses that trigger resonant capture of abundant elements in Earth~\cite{IceCube:2016aga}. 
A more detailed investigation of captured DM abundances is deferred to future work. 
 
\subsection{Modifications to primordial density fluctuations}

New particles and their associated dynamics in our framework may also leave an imprint on cosmological observables. One aspect is structure formation. For instance, the metastable $\chi_2$ can induce an EMD epoch, during which the subhorizon density perturbations grow linearly. In contrast, the motion of DM particles leads to damping of density perturbations at small scales.
As studied in the literature (e.g.,~\cite{Erickcek:2011us} and most recently in \cite{Delos:2021rqs}), the total effect on density perturbations is mainly decided by two scales: the horizon size at the end of the EMD epoch and the scale below which primordial fluctuations are suppressed by DM streaming out of over- and under-dense regions. %
 
The scale associated with perturbation growth is decided by the horizon size at cosmic time $t = \tau$, where $\chi_2$ decays and the EMD epoch ends. Its value can be expressed in terms of co-moving distance as
\begin{equation}\label{eq:lmax}
	L_{\rm end}= { \tau }\, \left( {s(T_{\rm end}) \over s (T_0) } \right)^{1/3} \approx 10^{-5}\,\sqrt{\tau \over 0.02\,{\rm s} }\,{\rm Mpc}\,,
\end{equation}
where $s(T)$ is the radiation entropy density at temperature $T$, $T_{end}$ is the reheating temperature post the EMD epoch; the photon temperature at present is $T_0=2.35\times 10^{-4}$\,eV~\cite{planck}. Perturbations at scales below $L_{\rm end}$ thus grow linearly w.r.t.~the scale factor~$a$ during this EMD epoch~\cite{Dodelson:2003ft, Erickcek:2011us}. 

On the other hand, density perturbations of DM can be suppressed due to random energy transfers either via sound waves or traveling particles, referred to as acoustic or collision(-less) damping~\cite{Green:2003un, Loeb:2005pm, Jeong:2014gna}.
Since $\chi_1$ does not significantly scatter after its freeze-out, collisionless damping dominates in the suppression of structure. This free-streaming scale is estimated by 
\begin{equation}\label{eq:fsLength}
L_{\rm fs}= a_0	 \int^{t_{\rm eq}}_{0} dt {v(t) \over a (t) }  \approx  10^{-9} \, {{\rm TeV} \over m_{\chi}/\xi_i} \,{\rm Mpc}\, \,
\end{equation} 
where we have neglected the logarithmic dependence on the exact time when DM kinetically decouples. 
Finally, we note that the last equality of Eq.~\eqref{eq:fsLength} assumes a standard cosmology. In presence of a matter-dominated epoch, the ratio of $a_0/a(t)$ becomes larger than in standard cosmology at $t<\tau$ (compare solid and dashed lines in right panel of Fig.~\ref{Fig:SimpleEvo}), thus a ``stretching factor'' $\Delta^{1/3}$ needs to be multiplied in this case. 

Although there exist two competing effects as outlined above (enhancement and suppression), observationally, the most relevant scale is the larger one of $L_{\rm end}$ and $L_{\rm fs}$. In our framework, the scale governing additional perturbation growth, $L_{\rm end}$, is larger than the suppression scale, $L_{\rm fs}$, as the former is the horizon size when DM is already non-relativistic. As a result, we generally expect a peak in density perturbations between the two scales, leading to a potentially observable enhancement in the local subhalo abundance.
The ensuing altered predictions of the ionization history as well as of the halo-mass function in the late Universe can be probed with various strategies~\cite{Erkal2016TheNA, Chluba:2019kpb, Furugori:2020jqn, Lee:2020wfn,Vikaeus2021ConditionsFD, Delos:2021rqs}.  
In particular, future Pulsar Timing Arrays (PTA) will be able to constrain the abundance of local subhalos down to a halo mass of $10^{-10}\,M_\odot$, and thus have the sensitivity to probe effects on primordial perturbations at co-moving scales as small as $10^{-7}$\,Mpc~\cite{Ramani:2020hdo, Lee:2020wfn}. Such a scale corresponds to $ \tau \sim 10^{-6}$\,s, or, conversely, a reheating temperature of $T_{\rm end} \sim 300$\,MeV following the end of the EMD era.

\section{Conclusions}\label{sec:conclusion}

In this work, we identify a new mechanism for the joint generation of the baryon and DM abundances. DM $\chi_1$ makes a thermal freeze-out via annihilation into a lighter metastable dark partner $\chi_2$ through an overall dark number conserving process $\chi_1 \chi_1\to \chi_2 \chi_2$. The lighter state $\chi_2$ subsequently decays to SM quarks. Its chemical decoupling and CP- and B-violating interactions with SM ensure the fulfillment of the Sakharov conditions, while its interactions with $\chi_1$ ensure fulfillment of the relic density requirement by observations. The lifetime of $\chi_2$ is assumed such that its decay happens after $\chi_1$-$\chi_2$ freeze-out yet before primordial nucleosynthesis. By itself, this only requires small couplings to SM, and, in general the dark and observable sector temperatures, $T'$ and $T$, may differ. 

We present a novel analytical treatment for a two-state dark sector freeze-out and subsequent baryogenesis assuming $T'\leq T$ and for which $T'=T$ is contained as a special case. There are then two principal options. First, in the hierarchical scenario there is a mass hierarchy $m_{\chi_1}/m_{\chi_2}\gtrsim 10$ such that $\chi_2$ is relativistic during $\chi_1$ freeze-out. The relic abundance of DM, $\chi_1$, is established following the dark freeze-out, and depends on the annihilation cross-section and the ratio $T'/T$, resembling the prediction familiar from WIMPs.
The $\chi_2$ abundance prior to its decay is also fixed by the freeze-out, yet is insensitive to the annihilation cross-section. Consequently, the ensuing prediction of
the baryon asymmetry is predominantly determined by the CP asymmetry $\epsilon_{\rm CP}$ and the initial value of $T'/T$. In the nearly degenerate scenario ($\delta \ll 1$) both $\chi_1$ and $\chi_2$ freeze out non-relativistically. However, because of overall $\chi_1 + \chi_2$ number conservation in the dark sector, its freeze-out yield again essentially coincides with that of radiation such as in the previous case. Therefore, the baryon asymmetry likewise only depends on~$\epsilon_{\rm CP}$ and the temperature ratio.

In either scenario, $\chi_2$ may dominate the energy budget of the Universe before its decay, leading to an early matter dominated era. The associated entropy injection dilutes both baryon asymmetry and DM abundance. However, it leaves their relative proportion unchanged. In the case of $s$-wave annihilation, the prediction for $\Omega_B/\Omega_{\rm DM}$ only depends on the DM annihilation cross section---which may take on its usual thermal value $O({\rm pb})$---and on~$\epsilon_{\rm CP}$, once the initial temperature ratio is fixed; see Eq.~\eqref{eq:swave}.
We verify our analytical estimates by numerically solving the Boltzmann equations, and find excellent agreement for benchmark parameter points. 

We then realize our general ideas by introducing a UV complete model. Here, a heavy pseudoscalar~$A$ mediates the dominant, $s$-wave annihilation of the fermionic DM state~$\chi_1$, while a massive CP-even scalar mediator $S$ may play a more favorable role for detection prospects in DM direct detection experiments. The connection to the SM is made through Yukawa interactions between fermionic states $\chi_{2,3}$, an $SU(3)_c$ charged scalar~$\phi$, and SM quarks. These interactions mediate the tree-level B-violating decay of $\chi_2$ which---through its interference with the loop-induced decay by the intermediate state $\chi_3$---becomes CP-violating. 
The colored state $\phi$ can, e.g., be pair-produced through gluons and decay to four jets, and is currently constrained by the LHC data to be at least of multi-TeV mass. Pair production of $\chi_{1,2,3}$ would lead to missing energy signals and/or in the form of displaced vertices.
Some of these collider signatures can be within reach of the High-Luminosity LHC, while future high energy colliders are more promising for detection. The conventional DM direct detection signal from this model is strongly suppressed due to the decoupled nature of the dark sector.
Model-specific indirect signals of baryogenesis from DM capture and annihilation in the Earth and other stellar objects, exotic signatures at large volume neutrino experiments, and departures of a standard matter power spectrum warrant further study.

In summary, we identify a new generic mechanism connecting the DM and baryon abundances. We term it ``Dark freeze-out Cogenesis'' on account of a separate thermal dark sector evolution in the early Universe, and that both DM abundance and baryon asymmetry are seeded during the same freeze-out dynamics in a dark sector. Our mechanism works over a vast range of dark sector masses.
The price to pay is restrictions on the size of the interactions with the SM that ensure certain decoupling conditions to hold. This renders direct and indirect tests more challenging, while these restrictions may be lifted with future model-building efforts in this direction.

\section*{Acknowledgments}
YC and JP thank KITP (supported by National Science Foundation under Grant No. NSF PHY-1748958) for hospitality.
XC is supported by the Austrian Science Fund FWF under Grant No.~FG1.
YC is supported in part by the US Department of Energy under award number DE-SC0008541.
 MS is supported by TRIUMF who receives federal funding via a contribution agreement with the National Research Council of Canada.
\appendix

\section{Freeze-out Solution in the Nearly Degenerate Scenario}
\label{app:Boltzmann}

The Boltzmann equation for the $\chi_1$ evolution is determined by $\chi_1$ annihilating into $\chi_2$. ``Inverse" annihilation of the lighter $\chi_2$ to heavier $\chi_1$ is forbidden at zero temperature but proceeds off the exponential tail of the $\chi_2$ velocity distribution. Assuming only this process for $\chi_1$ and $\chi_2$, the evolution of $\chi_1$ follows
\begin{equation}\label{eq:chi2ev}
\dot{n}_{\chi_1}+3Hn_{\chi_1}=-\langle\sigma_{\chi_1}v\rangle n_{\chi_1}^2+\langle\sigma_{\chi_2}v\rangle n_{\chi_2}^2 .
\end{equation}
The principle of detailed balance dictates that the right hand side of Eq.~\eqref{eq:chi2ev} vanishes in equilibrium, which gives the thermally averaged cross section for $\chi_2$ annihilation, $\langle\sigma_{\chi_2}v\rangle$, in terms of $\langle\sigma_{\chi_1}v\rangle$,
\begin{align}\label{eq:chi2ann}
\langle\sigma_{\chi_2}v\rangle = \bigg(\frac{n_{\chi_1}^\text{eq}}{n_{\chi_2}^\text{eq}}\bigg)^2\langle\sigma_{\chi_1}v\rangle =\bigg(\frac{g_{\chi_1}}{g_{\chi_2}}\bigg)^2(1-\delta)^{-3} e^{-2\delta x^\prime}\langle\sigma_{\chi_1}v\rangle \,,
\end{align}
where $\delta\equiv (m_{\chi_1}-m_{\chi_2})/m_{\chi_1}$ is the dimensionless mass splitting, $g_{\chi_1,\chi_2}$ are the number of internal degrees of freedom in $\chi_1$/$\chi_2$, and $x^\prime={m_{\chi_1}}/{T^\prime}$. The exponential factor arises from the equilibrium number density with a vanishing chemical potential, $n^\text{eq}=g\big(\frac{mT'}{2\pi}\big)^{3/2}e^{-m/T'}$. Plugging Eq.~\eqref{eq:chi2ann} into Eq.~\eqref{eq:chi2ev} and replacing the number densities $n_{\chi_{1,2}}$ with the co-moving densities $Y_{\chi_{1,2}}$ and cosmic time $t$ in favor of $x=m_{\chi_1}/T =x' \xi $, the $\chi_1$ abundance, $Y_{\chi_1} \equiv n_{\chi_1}(T')/s(T) $, evolves as 
\begin{equation}\label{eq:chi2coev}
\frac{dY_{\chi_1}}{dx}=-\frac{\lambda \xi^n}{x^{2+n}}\bigg[Y_{\chi_1}^2-(1-\delta)^{-3}e^{-2\delta x^\prime}Y_{\chi_2}^2\bigg]\,,
\end{equation}
with other quantities explained in the main text.
\begin{figure}%
\centering
\includegraphics[width=0.5\textwidth]{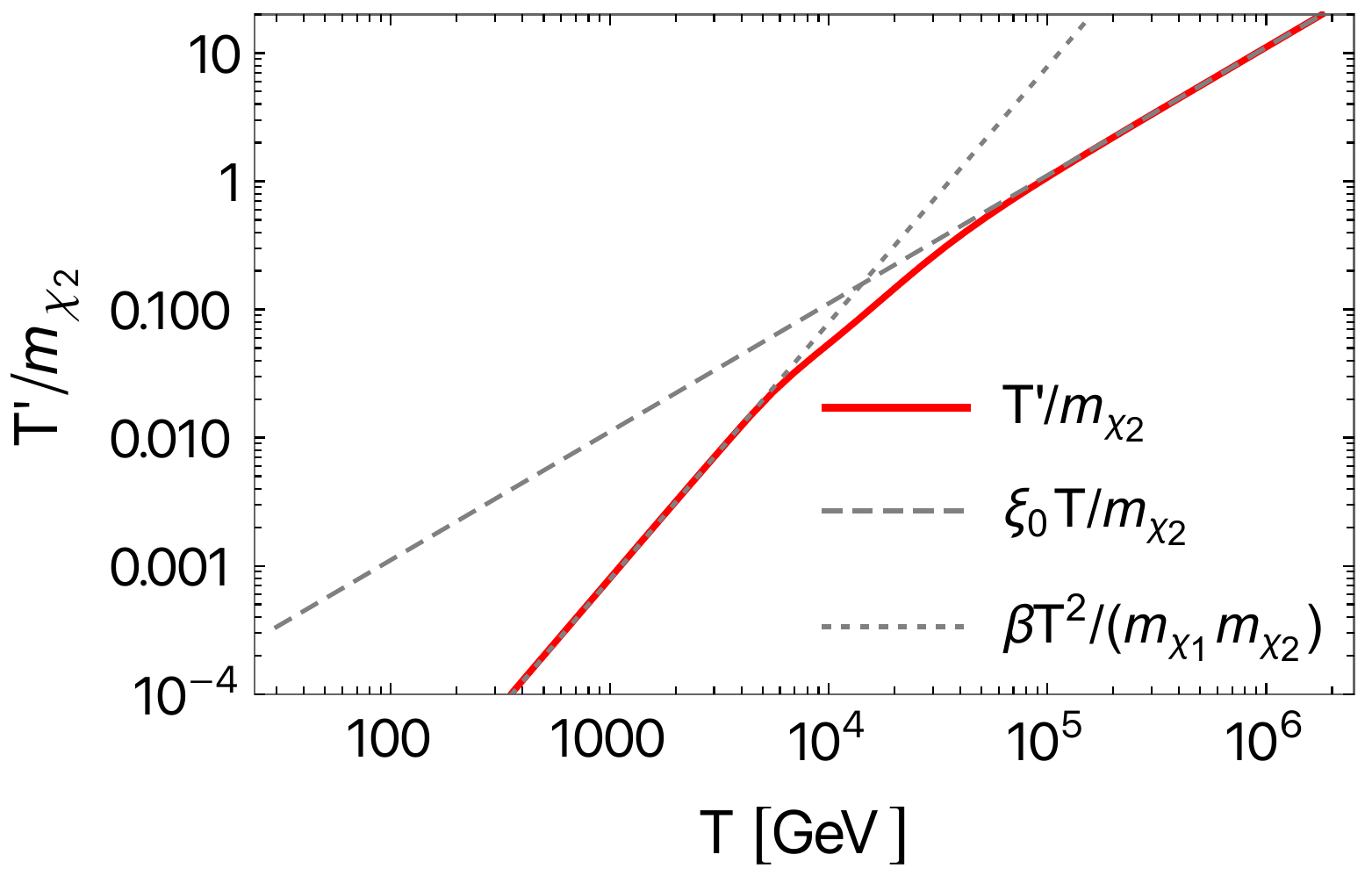}~
\includegraphics[width=0.5\textwidth]{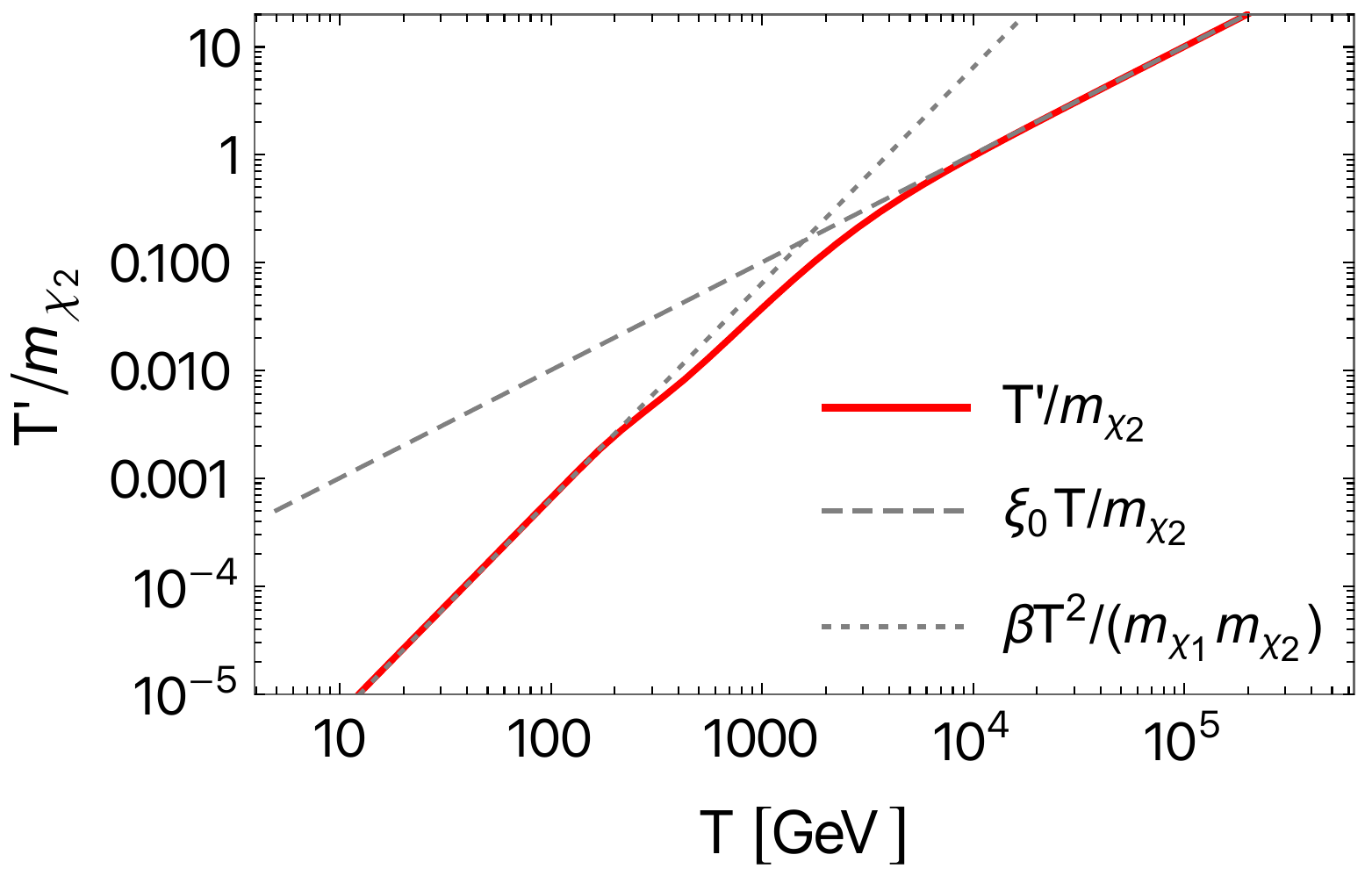}
\caption{Evolution of dark temperature $T'$ in the nearly degenerate scenario, where the couplings are fixed   to $g_1 g_2/m_A^2 =1/(4\,\text{TeV})^2$ and $m_{\chi_1} = 1$\,TeV, while the left panel takes  $\delta =0.1$, $\xi_i = 0.01$, and the right takes  $\delta =0.01$, $\xi_i = 0.1$. The red lines give exact values of $T'$, normalized to $m_{\chi_2}$, from numerically solving the Boltzmann equations. In contrast, gray dotted lines give a simple $T' \propto T$ scaling from its initial value at $T' \gg m_{\chi_2}$ , while gray dashed lines show the values from $(m_{\chi_1}/T') =\beta (m_{\chi_1}/T)^2$ of Eq.~\eqref{eq:empicialT}. It verifies that Eq.~\eqref{eq:empicialT}, with the scaling $T' \propto T^2$, works well in the regime of $T' \ll m_{\chi_2}$. 
}
\label{Fig:DarkTapp}
\end{figure}

At $T' \ll m_{\chi_1}$, number conservation requires $Y_{\chi_1} +Y_{\chi_2} \equiv  Y_{\chi_1,i}+ Y_{\chi_2,i} \approx 0.42g_{\chi}/g_{*S}\xi^3_i$, if setting initially the temperature ratio $\xi = \xi_i$ at $T' \ge m_{\chi_1}$. When the dark sector is in thermal equilibrium, the quasi-static condition, ${dY_{\chi_1}}/{dx} \approx 0$, needs to be satisfied for Eq.~\eqref{eq:chi2coev}, leading to $Y_{\chi_1} \approx (1-\delta)^{-3/2}e^{-\delta x^\prime}Y_{\chi_2}= (1-\delta)^{-3/2}e^{-\delta x^\prime}(Y_{\chi_1,i}+ Y_{\chi_2,i}-Y_{\chi_1}) $. In other words, the quasi-static equilibrium (QSE) solution of $Y_{\chi_1}$, defined as $Y_{\rm QSE}$, is expressed in terms of 
\begin{equation}\label{eq:QSE}
	Y_{\chi_1} \approx Y_{\rm QSE} \equiv \frac{(1-\delta)^{-3/2}e^{-\delta x^\prime} }{1+(1-\delta)^{-3/2}e^{-\delta x^\prime}}\times (Y_{\chi_1,i}+ Y_{\chi_2,i})\,.
\end{equation}
In general, the denominator is approximately one for cold freeze-out, i.e., if $x'_\text{f.o.} \gg 1/\delta$. We emphasize that our QSE solution is significantly different from that of a WIMP freeze-out. In the WIMP case, DM pair annihilate to final states that remain in chemical equilibrium with the thermal bath so that the associated Boltzmann equation is directly $dY_{\rm WIMP}/dx = -\lambda /x^{2+n}[Y_{\rm WIMP}^2 - (Y^{\rm eq}_{\rm WIMP} (\mu =0))^2]$, leading to a QSE solution $Y_{\rm QSE} = Y^{\rm eq}_{\rm WIMP} (\mu =0)$, when all relevant processes are sufficient. In our framework, non-vanishing chemical potentials arise for both dark particles due to the total number conservation of $\chi_1$ and $\chi_2$, which is why now its QSE solution has a dependence on the initial total dark abundance.

An implicit assumption here is that both particles are non-relativistic when freeze-out happens, i.e., our results below are not valid for $\delta\to 1$. The latter case is instead studied in the main text in the hierarchical scenario of Sec.~\ref{sec:scenario1}. 

To solve the equation \eqref{eq:chi2coev} at later time when the annihilation process gradually becomes out of equilibrium, we recast it in terms of $\Delta\equiv Y_{\chi_1}-Y_{\rm QSE} $ as
\begin{equation}\label{eq:depeq0}
\frac{d\Delta}{dx} \approx -\frac{dY_{\rm QSE}}{dx}-\frac{\xi^n \lambda }{x^{2+n}} \Delta \bigg[ \Delta + 2 Y_{\rm QSE}\bigg]\,,
\end{equation}
neglecting sub-leading terms under the assumption $\delta x'\gg 1$. Note that $\xi=T'/T$ is not a constant due to the fact that for a free non-relativistic particle $T' \propto 1/a^2$ while $T \propto 1/a$ for the radiative thermal bath. That is, one can set $\xi = 1/(\beta x)$, and thus $x'=\beta x^2$ in the non-relativistic limit $T' \ll m_{\chi_2}$. As argued in the main text, a reasonable choice for the coefficient is $\beta = (1-\delta)/(4\times 2^{2/3}\xi_i^2)$, as confirmed by the numerical results in Fig.~\ref{Fig:DarkTapp}. 
Using this, we may write
\begin{equation}\label{eq:depeq1}
\frac{dY_{\rm QSE}}{dx}  =- \frac{(Y_{\rm QSE})^2}{Y_{\chi_1,i}+ Y_{\chi_2,i}} \, \bigg[ 2\delta \beta x\, (1-\delta)^{3/2} e^{\delta \beta x^2}\bigg]  \,.
\end{equation}

Following standard procedure, we assume the L.H.S. of Eq.~\eqref{eq:depeq0} to vanish at freeze-out $x_\text{f.o.}$, and take ${\Delta}(x_\text{f.o.}) = c Y_{\rm QSE}(x_\text{f.o.})$, to solve for $x_\text{f.o.}$. As fiducial value, we choose $c =0.3$ with a rather mild dependence on the variation of that number. It then follows that at $x=x_\text{f.o.}$ we may write,
\begin{equation}
\frac{\xi^n \lambda }{x^{2+n}} c ( 2 + c ) = {-\frac{1}{(Y_{\rm QSE})^2}\frac{dY_{\rm QSE}}{dx}}  = \frac{1}{Y_{\chi_1,i}+ Y_{\chi_2,i}} \, \bigg[ 2\delta \beta x\, (1-\delta)^{3/2} e^{\delta \beta x^2}\bigg] \, .
\end{equation}
 Substituting $Y_{\chi_1,i}+ Y_{\chi_2,i} \approx 0.42g_{\chi}\xi^3_i/g_{*S}$ and $\xi = 1/(\beta x)$, the exponential above is expressed as
\begin{equation}\label{eq:Lf} 
e^{\delta \beta x_\text{f.o.}^2} = \frac{ \lambda }{x^{2+2n} \beta^n }c(2+c)\, { 0.42g_{\chi}\xi^3_i \over g_{*S}} {1 \over 2\delta \beta x\, (1-\delta)^{3/2} } = \frac{ 0.21 c(2+c) \lambda \xi^3_i \beta^{1/2} \delta^{1/2} }{(1-\delta)^{3/2} } \, {g_{\chi} \over g_{*S}} \left( {1\over \delta \beta x_\text{f.o.}^2} \right)^{n+3/2} \,, \notag
\end{equation}
where the L.H.S.~increases and the R.H.S.~decreases with growing $x_\text{f.o.}$, so that there is a unique solution to~$x_\text{f.o.}$. Taking the logarithm of both sides, and iteratively replacing $x_\text{f.o.}$ on the R.H.S.~give us 
\begin{equation}\label{eq:Lfo}
\delta \beta x_\text{f.o.}^2 =  \ln \left[ \frac{ 0.21 c(2+c) \lambda \xi^3_i \beta^{1/2} \delta^{1/2} }{(1-\delta)^{3/2} } \, {g_{\chi} \over g_{*S}} \right] - \left(n+{3\over 2}\right) \ln \left\{\ln \left[ \frac{ 0.21 c(2+c) \lambda \xi^3_i \beta^{1/2} \delta^{1/2} }{(1-\delta)^{3/2} } \, {g_{\chi} \over g_{*S}} \right] \right\}\,,
\end{equation}
assuming the first term dominates, to yield the observed DM relic abundance. 

\begin{figure}%
\centering
\includegraphics[width=0.49\textwidth]{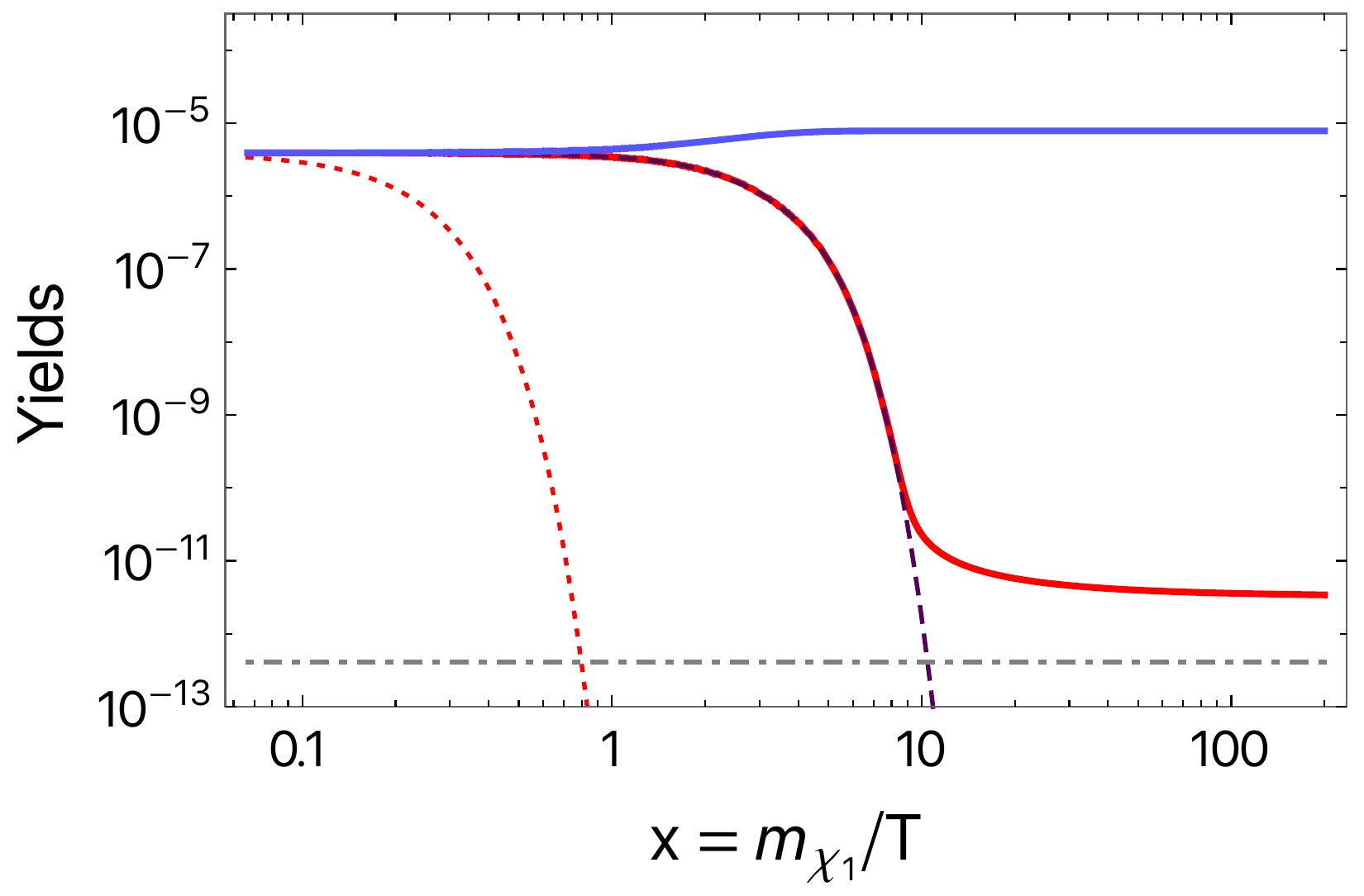}~
\includegraphics[width=0.49\textwidth]{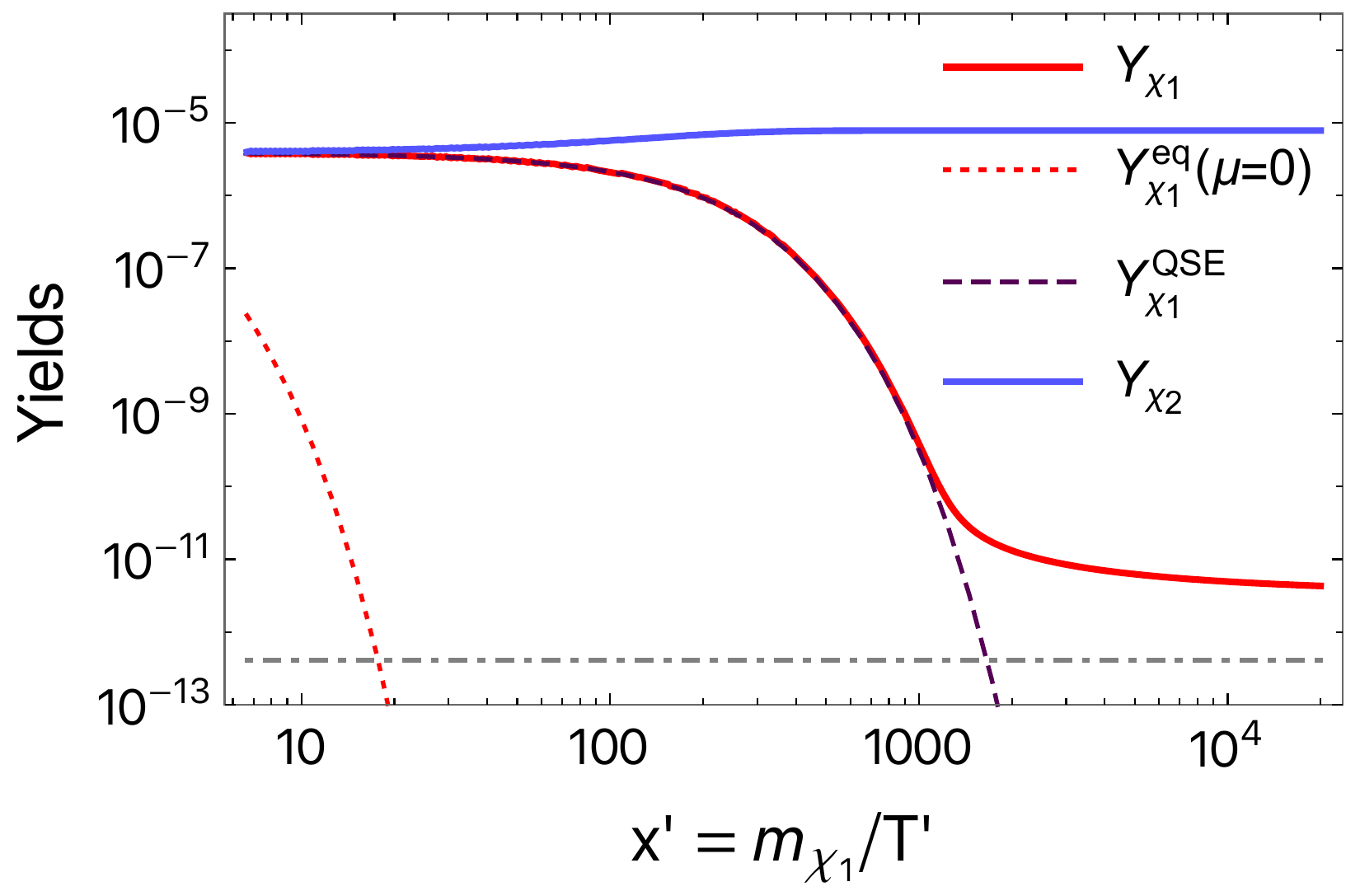}\\
\includegraphics[width=0.49\textwidth]{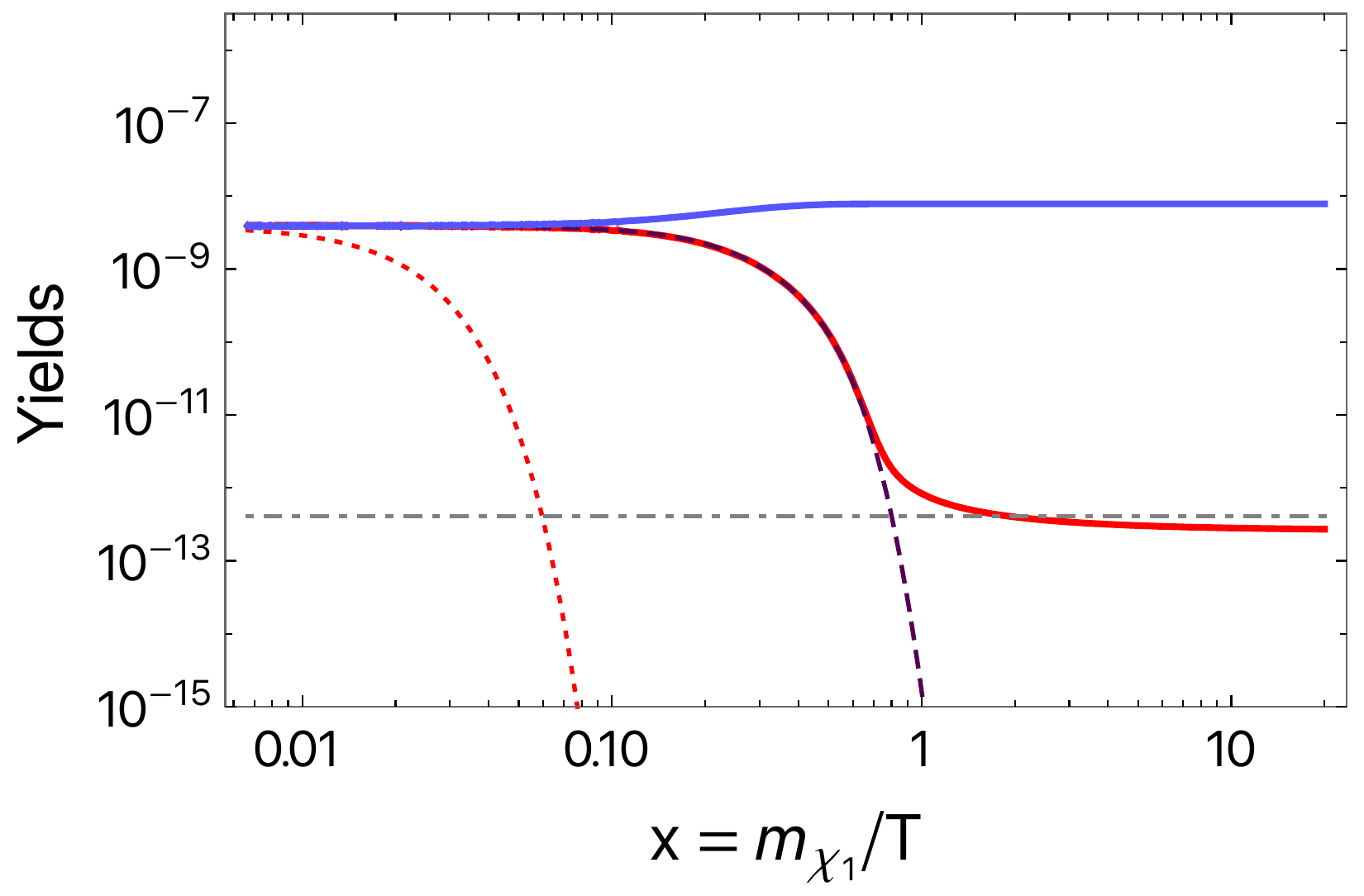}~
\includegraphics[width=0.49\textwidth]{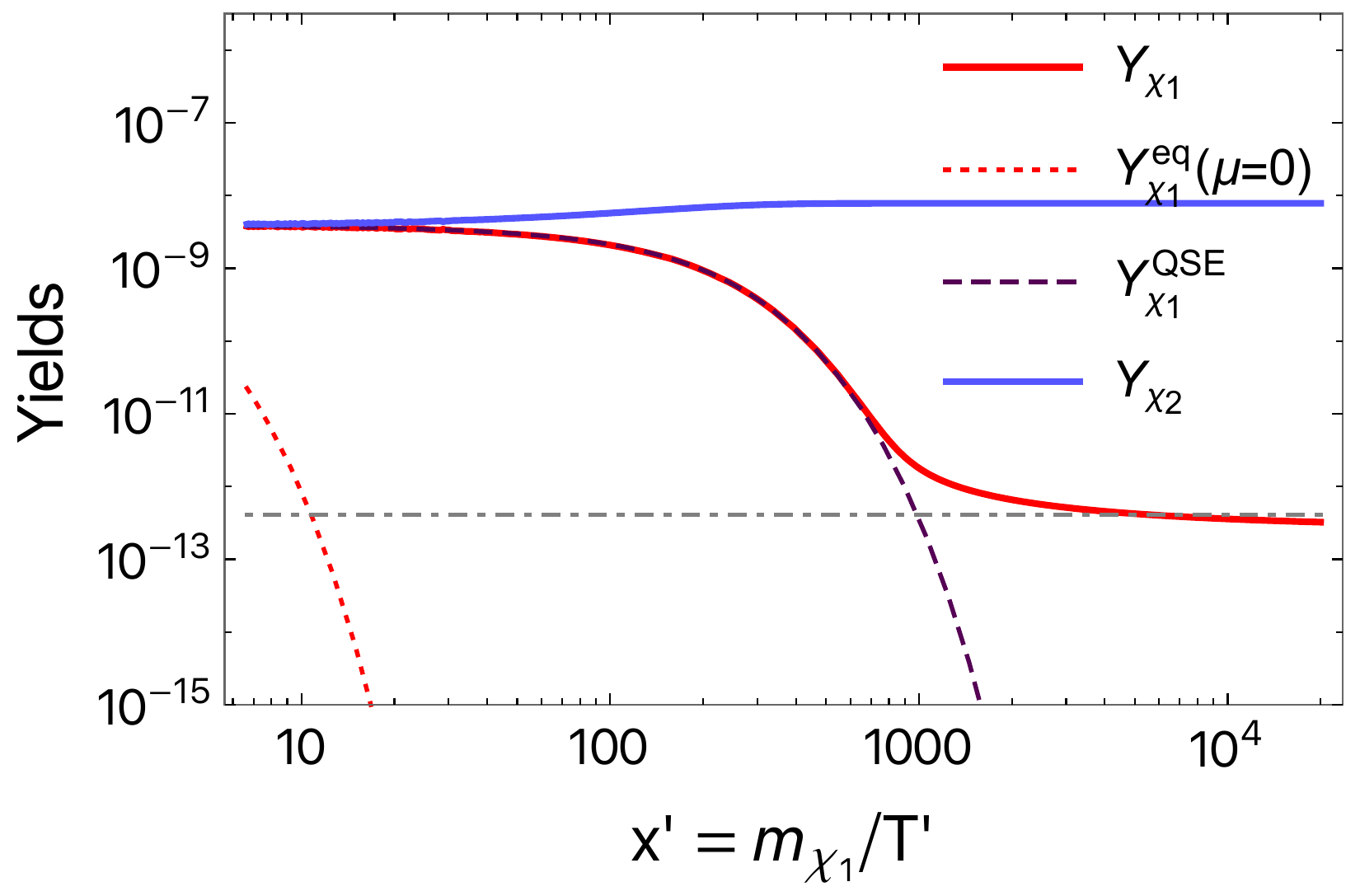}\\
\includegraphics[width=0.49\textwidth]{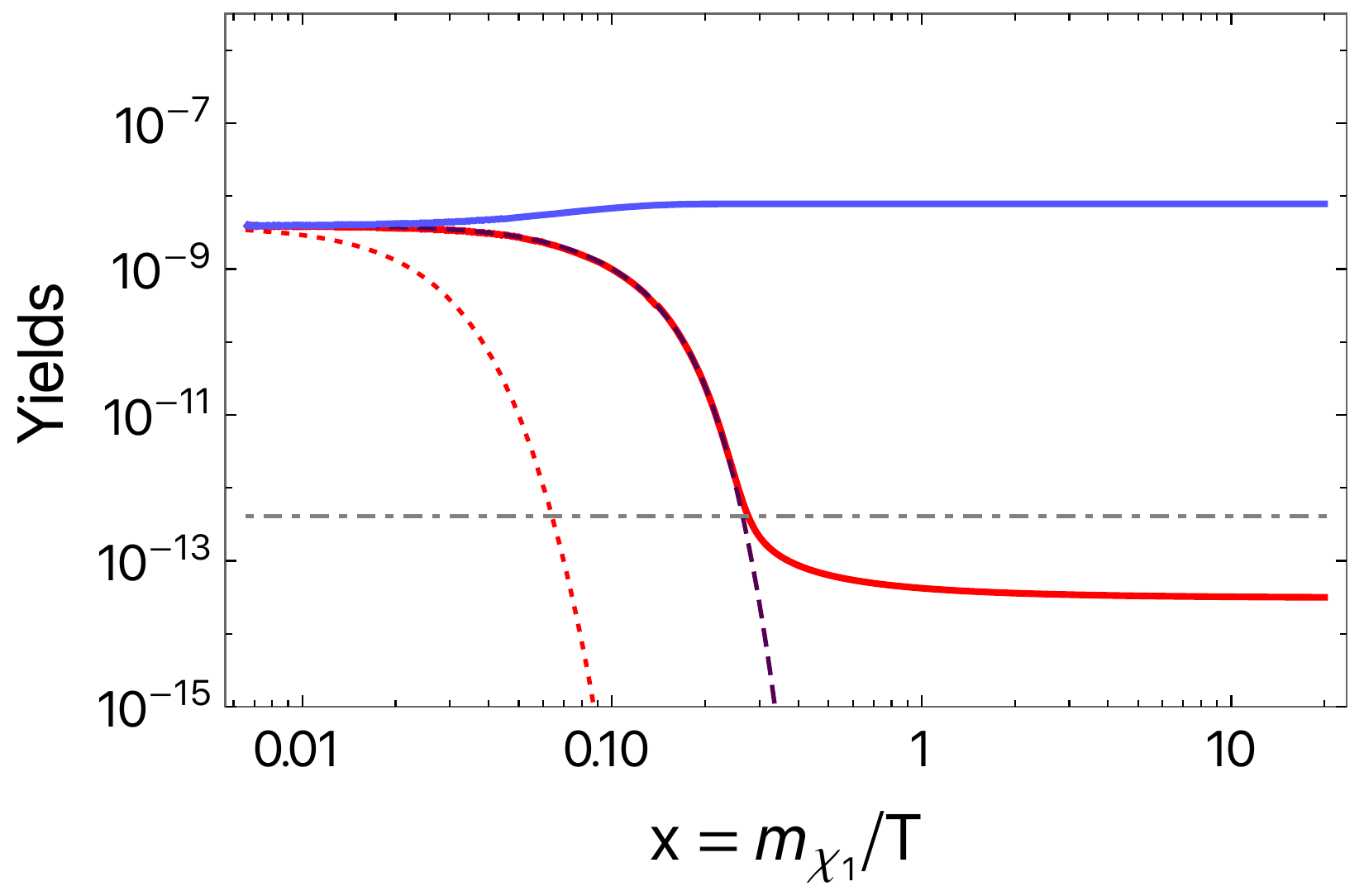}~
\includegraphics[width=0.49\textwidth]{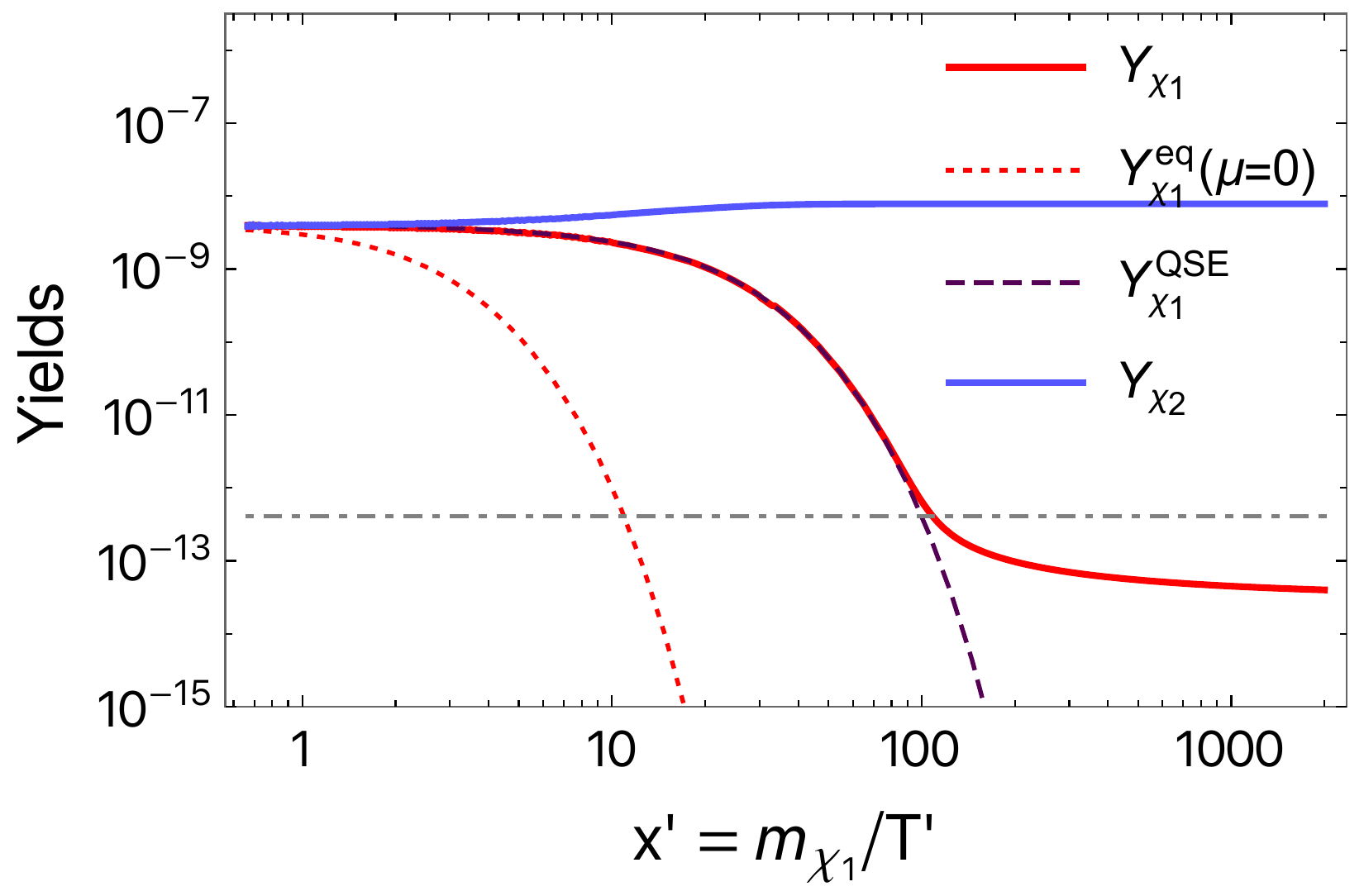}
\caption{The abundance evolution of $\chi_1$ and $\chi_2$ for $m_{\chi_1} = 1$\,TeV and $g_1 g_2/m_A^2 =1/(4\,\text{TeV})^2$ as a function of $x$ (left) and $x'$ (right), same as Fig.~\ref{Fig:Split01}, except that the approximation lines, $e^{-\delta x'}Y_{\rm tot}$, are omitted here. The additional parameters are modified to  $\delta =0.01$, $\xi_i =0.1$ (top),  $\delta =0.01$, $\xi_i =0.01$ (middle), and $\delta =0.1$, $\xi_i =0.01$ (bottom). }
\label{Fig:SplitAdd}
\end{figure}

Finally, under the assumption that $Y_{\rm QSE}( x_\text{f.o.}) \gg Y_{\chi_1,f} \gg Y_{\rm QSE}|_{x\to \infty} $, we obtain
\begin{equation}
\frac{d\Delta}{dx} \approx - \frac{\xi^n \lambda }{x^{n+2}}\Delta^2 = - \frac{ \lambda}{ \beta^n x^{2n+2} }\Delta^2  \,, 
\end{equation}
and integrating it over $x$ from $x_\text{f.o.}$ to $\infty$ results in 
\begin{equation}\label{eq:chi1abund}
	Y_{\chi_1, f} \approx \Delta|_{x\to \infty }\approx \frac{(2n+1)\beta^n}{ \lambda }x_\text{f.o.}^{2n+1}  = \frac{(2n+1) }{ \lambda \beta^{1/2} \delta^{n+1/2}} \left(\delta \beta x_\text{f.o.}^2\right)^{n+1/2}   \,, 
\end{equation}
where the last term is obtained from Eq.~\eqref{eq:Lfo}. We may verify our analytical results above by comparing to the numerical results shown in Fig.~\ref{Fig:SplitAdd}, where one finds good agreement. As expected, the evolution of $T'$ is insensitive to the detailed processes. This is because it is essentially governed by the dominant number abundance $Y_{\chi_2}$, and, thus, by the lighter state $m_{\chi_2}$ and~$\xi_i$.

\bibliographystyle{utcaps_mod}
\bibliography{ref}

\end{document}